\documentclass{agujournal2019}
\usepackage{url} 
\usepackage{ragged2e}
\usepackage{graphicx}
\usepackage{soul}
\usepackage{dirtytalk}
\journalname{JGR: Space Physics}

\begin{document}

%
%

\title{Predicting the Time-of-Arrival of Coronal Mass Ejections at Earth From Heliospheric Imaging Observations}

\authors{Carlos Roberto Braga \affil{1,2}, Angelos Vourlidas \affil {2}, Guillermo Stenborg \affil{3}, Alisson Dal Lago  \affil{4}, Rafael Rodrigues Souza de Mendon\c{c}a \affil{4}, Ezequiel Echer \affil{4}}

\affiliation{1}{George Mason University, 4400 University Drive, Fairfax, VA, 22030, USA}
\affiliation{2}{The Johns Hopkins University Applied Physics Laboratory, Laurel, MD 20723, USA}
\affiliation{3}{Space Science Division, U.S. Naval Research Laboratory, 4555 Overlook Ave. SW Washington, DC 20375, USA}
\affiliation{4}{National Institute for Space Research, Av. dos Astronautas, 1758, S\~{a}o Jos\'{e} dos Campos, 12227-010, SP, Brazil}

\correspondingauthor{Carlos Roberto Braga}{cbraga@gmu.edu}

\begin{keypoints}
\item We study CME propagation relying on simultaneous observations of Earth-directed CMEs from the inner Heliospheric Imagers onboard STEREO 

\item We adopt an elliptical front-fitting approach to the two HI-1 viewpoints and use a drag model to simulate the CME propagation in the heliosphere

\item We derive a CME Time-of-Arrival and Speed-on-Arrival mean absolute errors of $6.9\pm3.9$ hours and $117\pm102$ km/s for a set of 14 events

\end{keypoints}

\justify

\begin{abstract}
 The Time-of-Arrival (ToA) of coronal mass ejections (CME) at Earth is a key parameter due to the space weather phenomena associated with the CME arrival, such as intense geomagnetic storms. Despite the incremental use of new instrumentation and the development of novel methodologies, ToA estimated errors remain above 10 hours on average. Here, we investigate the prediction of the ToA of CMEs using observations from heliospheric imagers, i.e., from heliocentric distances higher than those covered by the existent coronagraphs. In order to perform this work we analyse 14 CMEs observed by the heliospheric imagers HI-1 onboard the twin STEREO spacecraft to determine their front location and speed. The kinematic parameters are derived with a new technique based on the Elliptical Conversion (ElCon) method, which uses simultaneous observations from the two viewpoints from STEREO. Outside the field of view of the instruments, we assume that the dynamics of the CME evolution is controlled by aerodynamic drag, i.e., a force resulting from the interaction with particles from the background solar wind. To model the drag force we use a physical model that allows us to derive its parameters without the need to rely on drag coefficients derived empirically. We found a CME ToA mean error of $1.6\pm8.0$ hours ToA and a mean absolute error of $6.9\pm3.9$ hours for a set of 14 events. The results suggest that observations from HI-1  lead to estimates with similar errors to observations from coronagraphs.
\end{abstract}

 \section{Introduction}
 \label{sec:intro}
Coronal mass ejections (CMEs) have been tracked with space-based coronagraphs for more than 40 years. Thousands of events have been studied and catalogued  \cite{Tousey1973, Gosling1974, Howard1985, Webb1994, Gopalswamy2004, Robbrecht2004, Yashiro2004,Gopalswamy2009, Gopalswamy2016, Vourlidas2017, Lamy2019}. One of the key open issues about CMEs is understanding their propagation in the heliosphere, especially for events directed to Earth.

CMEs are the main drivers of intense geomagnetic storms \cite{Gosling1993} and one of the most basic variables from a Space Weather perspective is the Time-of-Arrival (ToA) of a given CME in the Earth's vicinity. Not surprisingly, the ToA has been studied for a long time. An extensive review of methods to estimate the ToA and their results can be found in \citeA{Zhao2014} and \citeA{Vourlidas2019}. 

The methods applied to ToA estimation include empirical approaches \cite{Gopalswamy2001,Schwenn2005, Kilpua2012, Makeda2016, Mostl2017}, magneto-hydrodynamic (MHD) modeling \cite{Wold2018, Mays2015}, CME three dimensional (3D) reconstruction, and CME propagation analysis based on drag-based models \cite{Vrsnak2014,Shi2015, Napoletano2018}, just to name a few \cite<see, e.g.,>{Zhao2014}. In spite of the insight gained with the dual view-point provided by the Solar Terrestrial Relations Observatory (STEREO) mission \cite{Kaiser2007} since 2007, the uncertainty of the CME ToA persist.  According to a review from \citeA{Vourlidas2019}, the CME ToA mean absolute error (MAE)  is $9.8\pm 2$ hours. This estimate is based on a comprehensive sample of ToA published studies, which make use of several, distinct methodologies to derive the CME ToA. The identification of the source of these errors is not straightforward. The evaluation and comparison of the different methodologies across the literature is complicated because of the different event samples, assumptions with regard to the propagation of the events in the interplanetary medium (i.e., beyond the field of view of the instrument utilized), and ToA criteria. Hence, it is difficult to asses which methodology yields the best results, i.e., the lowest ToA error. In general, the review from \citeA{Vourlidas2019} indicates that the errors tend to be lower for small sample studies, which is most likely due to an event-selection bias.

Another basic CME impact parameter for Space Weather is the Speed-on-Arrival (SoA). To derive this parameter observationally, the CME kinematics are usually derived in the corona and extrapolated to 1 au, normally using empirical or physics-based models. Then, by comparing the estimated SoA with the CME speed measured in situ, one can determine the SoA error. This error has been investigated in a limited number of studies in the STEREO era. This is evident in a recent review from \citeA{Vourlidas2019}. Among 24 studies including ToA error analysis, only 5 included SoA errors. Now, we briefly introduce these 5 studies along with their median SoA errors. Using several methods to determine the front, such as fixed-phi, harmonic mean and self-similar expansion fitting (described latter in this section) for a set of 22 CMEs, \citeA{Mostl2014} found median SoA errors in the 200-300 km/s range, depending on the method used. \citeA{Corona-Romero2017} estimated the SoA using a theoretical piston shock model combined with an empirical relationship for 40 fast CMEs, which are typically preceded by shock waves. The median SoA error they found is $95\pm249\ km/s$. Probably the most extensive study including SoA errors is \citeA{Mostl2017}, which includes more than 50 events in each STEREO viewpoint. The median SoA error found using self-similar expansion fitting is $191\pm341\ km/s$ for STEREO-A events and $245\pm446\ km/s$ for STEREO-B. Other studies cover fewer CME events and find smaller errors. \citeA{Hess2015} used a flux rope geometrical model \cite{Thernisien2006b,Thernisien2011} for the CME front and a prolate spheroid bubble model \cite{Kwon2014} for the sheath front associated with the CME. Combining both models with the drag force, the average SoA error found was $24.5\ km/s$ for their set of 7 events. One of the lowest SoA median errors was found by \citeA{Rollett2016} using an elliptical CME front model: for a set of 21 events, the median error is lower than 20 km/s. 

Thanks to the Sun-Earth Connection Coronal and Heliospheric Investigation (SECCHI) suite onboard STEREO, CMEs can be observed further into the inner heliosphere by the heliospheric imagers (HI-1 and HI-2), typically up to heliocentric distances between $0.5\ au$ and $1\ au$. Details about SECCHI are described in \citeA{Howard2008}.
Nevertheless, a quick check of the literature reveals that the use of the imaging products of the heliospheric imagers is limited compared to those of the coronagraphs \cite{Vourlidas2019,Zhao2014,Harrison2017}. 

In the heliospheric imagers FOV, the position of the CME can  be derived only under assumptions about the CME trajectory. Widely-used methods include the Fixed-$\phi$ (f-$\phi$) \cite{Sheeley1999,Kahler2007, Rouillard2008} and harmonic mean (HM) methodology \cite{Lugaz2009}. The former considers the CME as a point-like structure moving radially away from the Sun with constant speed  to determine the direction of propagation and CME position. The HM considers a circular structure centered in the Sun with half-width of $90^{\circ}$ propagating at constant speed. A third method is the self-similar expansion fitting (SSEF), which also assumes a circular front with half-width constant over time but adjustable to each CME \cite{Lugaz2010, Davies2012, Mostl2013}.

The f-$\phi$, HM and SSEFs methodologies allow us to determine the CME front position using a single viewpoint. These methods have been extensively applied to CMEs observed in the STEREO-era and the results are publicly available in the HELCATS (Heliospheric Cataloguing, Analysis and Techniques Service) project  (\url{www.helcats-fp7.eu/}). Many other catalogs are part of this project, such as the HELCATS Heliospheric Imager Geometrical Catalogue (HIGeoCAT), which reports kinematic properties derived using single-spacecraft observations of CMEs  observed by the HI-1 and HI-2 instruments, including their speeds, propagation directions, and launch times \cite{Mostl2017,Barnes2019}. Another list of CME kinematic parameters based on HI-1 observations is available in \url{http://www.stereo.rl.ac.uk/HIEventList.html}. Some studies also compare the results of multiple CME kinematics and ToA derived using the different methodologies mentioned above \cite{Mostl2014}.

Triangulation is yet another methodology to derive the CME kinematics. In this case, co-temporal observations are needed, e.g., from the twin heliospheric imagers or coronagraphs onboard STEREO \cite<e.g.,>{Liu2010, Liu2011, Liewer2011, Braga2017}. This methodology normally requires selection and tracking of particular point-like features in each viewpoint. It uses epipolar geometry, which allows the use of multiple viewpoints, and it normally requires assumptions about the structure under study.

To properly locate and track the CME fronts, and hence kinematically characterize the CME evolution in heliospheric images, further analysis is required (as compared to coronagraph observations) beyond the assumptions discussed above.
The relative contribution of the electron corona signal (i.e., the K-corona) to the total signal recorded by the HI instruments for elongations greater than about $8^\circ$ ($\sim 32\ R_\odot$) is well below that recorded by coronagraphs. 
Therefore, to help reveal the CME boundaries and inner structure during their evolution across the HI instrument field-of-view (FOV), it is necessary to remove the dominant signal coming from the F-corona, i.e., photospheric light scattered by the dust particles in orbit around the Sun \cite{Leinert1998}.

In addition, at the solar elongation covered by the heliospheric imagers, the emission properties of the coronal electrons change due to Thomson Scattering \cite{Minnaert1930}. The maximum brightness contribution along the line of sight is now located on the \say{curved} Thomson sphere rather than the flat \say{sky-plane} (\citeA{Vourlidas2006} and references therein). This effect complicates the visualization of the event boundaries, as CMEs move away from the Sun.

A motivation for this work is the application of a similar methodology to CME observations in the inner heliosphere from upcoming and planned missions, such as the recently selected PUNCH or L5-mission concepts \cite{vourlidas2015}. Future observations can be used in combination with a second spacecraft observing the same region, such as STEREO-A.

To carry out the investigation, we apply a customized version of the technique developed by \citeA{Stenborg2017} to remove the background signal in the HI-1 FOV on a set of 14 Earth-directed CME events spread over the rise and maximum of Cycle 24 (2010-2013). Co-temporal HI-1 observations from two viewpoints are used to construct an elliptical model of the CME fronts and hence estimate their locations in the solar corona. Beyond the HI-1 FOV, we apply a drag force model to propagate the CME up to 1 au. We finally compare the CME ToA errors computed with this approach to those calculated using mainly observations from SECCHI coronagraphs.

This article is organized as follows. In Section \ref{sec:list} we describe the events studied. From Section \ref{sec:removal} to Section \ref{sec:background_wind}, we describe the methodology applied to calculate the CME kinematics in the HI-1 FOV and extrapolate them in the remaining trajectory toward the Earth. The results (the calculated CME travel time, final speed, etc.) and a comparison with the actual observations are shown in Section \ref{sec:results}. Finally, we summarize the results in Section \ref{sec:conclusions}.

 \section{Materials and Methods}
 \label{sec:mat_met}
 
 We devised a methodology to estimate the CME ToA by combining a geometric front reconstruction model with a CME propagation model. To obtain the CME propagation direction, we fit the CME front in the ecliptic plane with an ellipse (see Section \ref{sec:elliptical}). To determine the CME kinematics, we use an aerodynamic drag force model (see Section \ref{sec:dragmodel}). The elliptical front allows us to estimate the initial position and the speed, which are then used as input parameters for the drag force estimation. As a final result, we derive the CME speed and ToA at $1\ au$. 

 \subsection{Event List}
\label{sec:list}
 Our starting point is the list of CME events analyzed by \citeA{Sachdeva2017}, which includes 38 well-observed events between March 2010 and March 2013.  This list includes only events with continuous observations in all STEREO SECCHI instruments, including the coronagraphs and heliospheric imagers, as well as from the Large Angle Spectrometric  Coronagraph \cite<LASCO; >{Brueckner1995} C2 instrument onboard the Solar and Heliopheric Observatory \cite<SOHO; >{Domingo1995}. Since our study targets only observations from heliospheric imagers, we do not use the kinematic parameters and height-time profiles derived by \citeA{Sachdeva2017} because they were obtained using observations from  coronagraphs (SECCHI and LASCO) and heliospheric imagers. We consider only the timing of each event in the list to identify the corresponding observations on the HI-1 FOV. Moreover, \citeA{Sachdeva2017} did not identify the CME counterparts in the Earth's vicinity (the so-called interplanetary coronal mass ejections - ICMEs); therefore, we undertake this task for each event.  
 
In order to perform this task, we use the ICME list compiled from WIND mission observations from \citeA{Nieves-Chinchilla2018}, which is available online at \url{https://wind.nasa.gov/ICMEindex.php}. Our criterion to associate a given ICME to its corresponding CME counterpart is based on the time elapsed ($t_{el}$) between the ICME in situ observation time and the time of the first coronagraph observation of the CME counterpart candidate. The CME travel time considered was taken from an extensive study of CME-ICMEs pairs by \citeA{Richardson2010}. We consider it a match when 0.5 days $< t_{el} <$ 5 days.  
 
 \begin{table}
 \centering 
 \caption{List of CMEs and corresponding ICMEs. The symbols $\|$ indicate the events removed from the list due to observation of another CMEs in close timing. Events associated to multiple ICMEs, without any ICME associated, or whose time elapsed between the CME and corresponding ICME observation falls outside our criteria are indicated by $\Pi$, $\emptyset$ and $\Delta$, respectively. We could not apply the F-corona background removal methodology (see Section \ref{sec:removal}) to events indicated by $\dag$ and they were removed from the analysis. The final list of 14 events that match all the criteria explained in Section \ref{sec:list} and that could be processed as explained in Section \ref{sec:removal} are indicated by a star.}
 \label{tab:events}
 \scalebox{0.7}{
\begin{tabular}{c c c c c c }
 \hline
ID	& Remark & Date	& Realistic timing  & Unique ICME &	ICME start date \\
 \hline
1 & $\dag$	        &2010/03/19	&yes		&yes	&2010/03/23 22:29\\
2 &$\star$	        &2010/04/03	&yes		&yes	&2010/04/05 07:55\\
3 &$\star$	        &2010/04/08	&yes		&yes	&2010/04/11 12:20\\
4 &$\|$             &2010/06/16	&yes		&yes	&2010/06/21 03:35\\
5 &$\dag$	        &2010/09/11	&yes		&yes	&2010/09/15 02:24\\
6 &$\dag$	        &2010/10/26	&yes		&yes	&2010/10/31 02:29\\
7 &$\emptyset$	    &2010/12/23	&-			&-		&-\\
8 &$\Delta$	        &2011/01/24	&Too short	&yes	&2011/01/24 06:43\\
9 &$\star$	        &2011/02/15	&yes		&yes	&2011/02/18 01:50\\
10&$\emptyset$      &2011/03/03	&-			&-		&-\\
11&$\star$	        &2011/03/25	&yes		&yes	&2011/03/29 15:12\\
12&$\emptyset$	    &2011/04/08	&-			&-		&-\\
13&$\star$	        &2011/06/14	&yes		&yes	&2011/06/17 02:09\\
14&$\emptyset$	    &2011/06/21	&-			&-		&-\\
15&$\emptyset$	    &2011/07/09	&-			&-		&-\\
16&$\emptyset$	    &2011/08/04	&-			&-		&-\\
17&$\star$	        &2011/09/13	&yes		&yes	&2011/09/17 02:57\\
18&$\|$   	        &2011/10/22	&yes		&yes	&2011/10/24 17:41\\
19&$\|$   	        &2011/10/26	&yes		&yes	&2011/11/01 08:09\\
20&$\|$   	        &2011/10/27	&yes		&yes	&2011/11/02 00:21\\
21&$\Pi$ 	        &2012/01/19	&yes		&no 	&2012/01/21 04:02\\
22&$\emptyset$	    &2012/01/23	&-			&-		&-\\
23&$\emptyset$	    &2012/01/27	&-			&-		&-\\
24&$\dag$           &2012/03/13	&yes		&yes	&2012/03/15 12:35\\
25&$\star$	        &2012/04/19	&yes		&yes	&2012/04/23 02:15\\
26&$\star$	        &2012/06/14	&yes		&yes	&2012/06/16 09:03\\
27&$\star$	        &2012/07/12	&yes		&yes	&2012/07/14 17:39\\
28&$\star$	        &2012/09/28	&yes		&yes	&2012/09/30 10:14\\
29&$\star$	        &2012/10/05	&yes		&yes	&2012/10/08 04:12\\
30&$\star$	        &2012/10/27	&yes		&yes	&2012/10/31 14:28\\
31&$\|$	            &2012/11/09	&yes		&yes	&2012/11/12 22:12\\
32&$\Delta$	        &2012/11/23	&Too short	&yes	&2012/11/23 20:51\\
33&$\star$	        &2013/03/15	&yes		&yes	&2013/03/17 05:21\\
34&$\star$	        &2013/04/11	&yes		&yes	&2013/04/13 22:13\\
35&$\Delta$	        &2013/06/28	&Too long	&yes	&2013/07/04 17:17\\
36&$\Pi$	        &2013/09/29	&yes		&no 	&2013/10/02 01:15\\
37&$\dag$           &2013/11/07	&yes		&yes	&2013/11/08 21:07\\
38&$\dag$           &2013/12/07	&yes		&yes	&2013/12/08 07:31\\
    \end{tabular}}
\end{table}

From our initial list comprising 38 events (Table~\ref{tab:events}) we could identify  30 ICME counterparts. The 8 unmatched events are $\#7$, $\#10$, $\#12$, $\#14$, $\#15$, $\#16$, $\#22$ and $\#23$ (indicated by $\emptyset$ in column \say{Remark}). In these cases, either the corresponding ICME was not included in the ICME list possibly due to data gaps or poor data quality, or the CME reported on \citeA{Sachdeva2017} missed the WIND spacecraft. 

In two other cases ($\#8$ and $\#32$, rows labeled \say{too short} in Table \ref{tab:events}), the ICME event arrives at Earth less than 24 hours after the time of first appearance in the coronagraph FOV as reported in \citeA{Sachdeva2017} in spite of the very low speed reported for this CME event. For other event ($\#35$, row labeled \say{too long}), the travel time is longer than 5 days and hence it is not considered to be a reasonable CME-ICME pair. Since we have not made a comprehensive study of all other CMEs observed in close timing to each of our events (few days before and after and including other instruments from the SECCHI suite or from LASCO coronagraphs C2 and C3), some of the three ICME with unreasonable travel times are perhaps not associated to the CMEs under study. We considered that these CME-ICME associations are inconsistent and removed them from our analysis. 

In two particular cases (events $\#21$ and $\#36$, indicated by $\Pi$ in column \say{Remark} of Table \ref{tab:events}), there is more than one ICME candidate. These events were also removed from our analysis because it is impossible to confirm (at least with the data we are using in the current analysis) which ICME corresponds to the CME arrival.

We also removed CME events from the list of \citeA{Sachdeva2017} that were preceded or followed closely (within less than 24 hours) by other CME events in the region close to the ecliptic plane. We performed this analysis on HI-1 FOV only (both on STEREO-A and STEREO-B) and considered the timing of the first observation of each event. These events are removed from the travel time analysis because interaction between consecutive CMEs is likely before their arrival at 1 au and after their observation on HI-1. Five events fit this criterion ($\#4$, $\#18$, $\#19$, $\#20$ and $\#31$) and are indicated by $\|$ in column \say{Remarks} of Table \ref{tab:events}.  When CME-CME interaction takes places, a detailed study would be necessary because additional forces need to be taken into account in the CME propagation to estimate their travel times (see, e.g., \citeA{Liu2012, Colaninno2013, Temmer2012}, just to mention some recent studies). Case studies of CME interaction are beyond the scope of the present manuscript. Notice that our criteria do not remove a given event from our list if CME interaction takes place below the HI-1 FOV or if one of the CMEs lies significantly northward or southward from the ecliptic plane. 

Therefore, from the original list of 38 events, 20 were left after the application of all criteria mentioned in the paragraphs above.  The final list studied here has 14 events because additional 6 events are eliminated when applying the methodology to remove the background F-corona, as explained in Section \ref{sec:removal}.

\protect It is worth mentioning here how our list of events compares to the total number of events available during the entire period of two-viewpoint observations from STEREO. Although hundreds of Earth-directed CMEs were observed in the period here selected \cite<see, e.g.,>{Barnes2019}, we expect that our strict selection criteria would eliminate the majority of them, hence reducing the number of events significantly. As an illustration, the number of ICMEs available from 2007 to 2014 is 138 according to the Wind ICME catalog. After removing ICMEs observed in close timing (one of our criteria requirements), 118 are left. Moreover, each of these ICMEs need to be associated to a unique CME observed on both HI-1 instruments (i.e, on STEREO-A and on STEREO-B). Continuous observations recording the passage of the event across the FOV in both instruments is the next requirement, along with the absence of another event prior or after the case under study in a time period of a few hours. A precise number of the remaining events would require an extensive case-by-case study to check our selection criteria, which is outside the scope of the present study. 

Briefly, from the list of 38 events used in the manuscript with corresponding ICMEs, we ended up with 14 events, i.e., the number of events is reduced to less than half after applying our selection criteria. Therefore, if we assume that the ratio of events selected to the sample size is kept, then the list of 138 ICMEs would have resulted in approximately 50 events. So, we estimate that the event list studied here corresponds to approximately one forth of the events that follow our criteria in the entire two-viewpoint STEREO period (2007-2014).

\subsection{Removal of the background F-corona}
\label{sec:removal}
The HI-1 observations include a background scene that must be removed to allow the CME event tracking and characterization. This background scene is dominated by the scattering from dust particles in orbit around the Sun, the so-called F-corona (the F letter stands for Fraunhofer). The F-corona intensity overtakes the K-corona above approximately $5R_\odot$ \cite{Koutchmy1985}, well below the inner edge of the HI-1 instrument, which is about $ 16R_\odot$. 

Experience from observations of the corona over the last 40 years suggests that the F-corona is  constant over  timescales of days or weeks while the K-corona is highly dynamic and can change significantly in a matter of hours. For this reason, empirical models of the F-corona are usually constructed by computing the minimum of the daily median images over an extended period of time (normally a solar rotation), centered on the day of observation \cite{Morrill2006}.

\citeA{Stenborg2017} showed that at the larger elongations covered by the HI-1 instruments, the use of background models obtained considering extended periods of time leads to the introduction of artifacts. This occurs due to the subtle changes resulting from different viewpoints \cite{Stenborg2018}. Therefore, to remove the background contribution from the F-corona from each individual HI-1 observation, we created its respective background model following \citeA{Stenborg2017}. 

\begin{figure}
 \noindent\includegraphics[width=\textwidth]{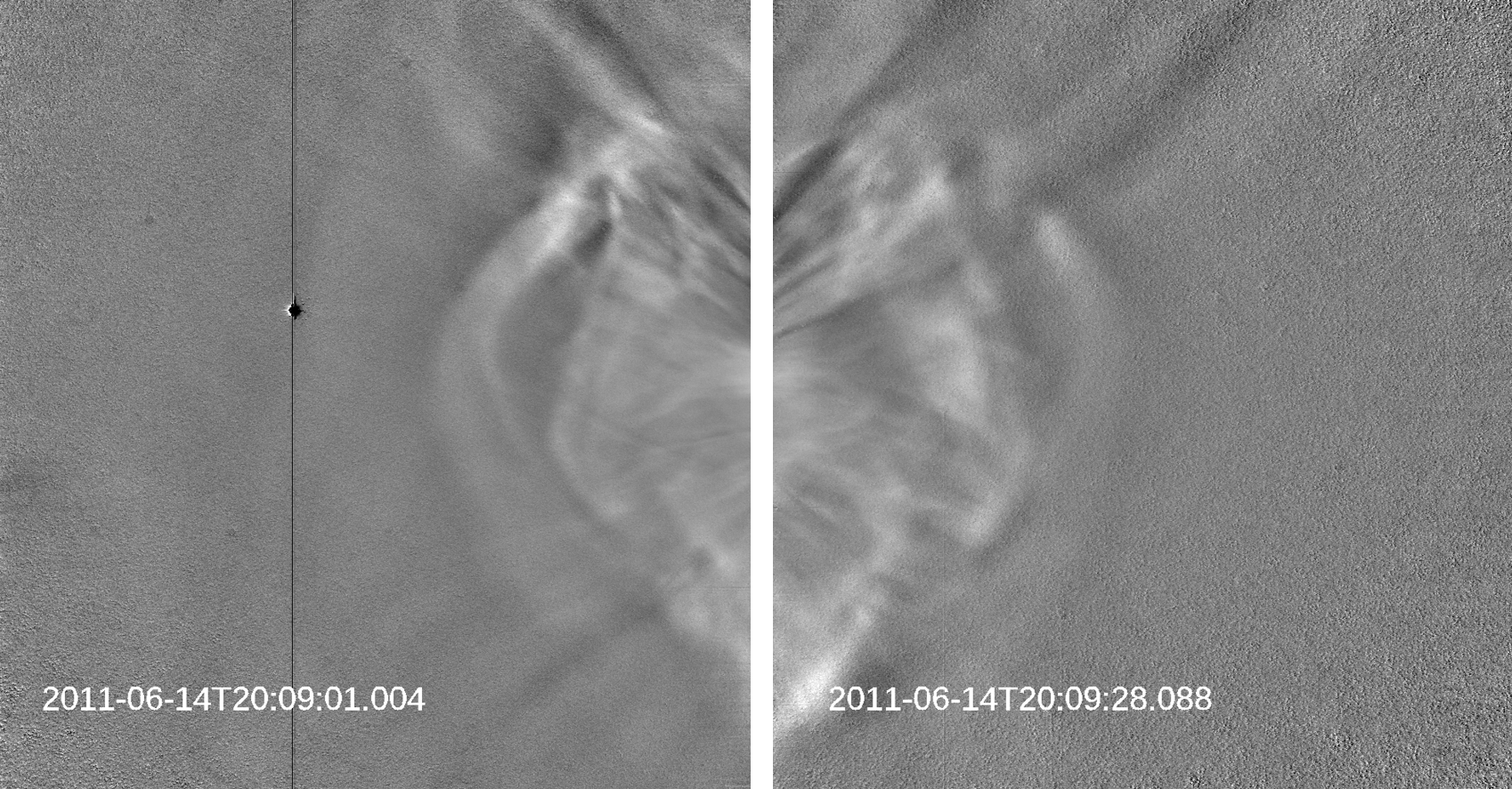}
\caption{Example of CME $\#13$ (2011/06/14) observed on HI-1-A (left) and HI-1-B (right) FOVs after removal of the background F-corona.}
\label{fig:hi1ab_example}
\end{figure}

An example of a processed HI-1 observation pair, highlighting CME feature, is shown in Figure \ref{fig:hi1ab_example}. The images reveal the often-seen (in coronagraphs) faint-bright front pair (shock or wave followed by the flux rope and a cavity, see \citeA{Vourlidas2013} for details), as well as more complex internal structure without the known artifacts that result from the use of the running difference scheme generally adopted by the solar physics community. In this scheme, each image is subtracted from a base image, which is typically taken a few time-steps behind. Background structures, such as streamers, that change over the course of hours, result in artifacts in the  final image. These artifacts normally prevent us from identifying internal structures of the CME such as core and void, and, depending or their size or relative brightness, the apex of a given CME \cite{Stenborg2017}.

In the current study, we focused on the selection of the CME furthermost point visible in the HI-1 FOV at each time-instance and at a position angle close to the ecliptic plane. Since we are interested in the arrival of the transient at the Earth, we did not take any measurement of their internal structure (e.g., the core of the events) but we also did not differentiate between shock and CME front, which may add some error in our ToA estimates.

As mentioned above, for some events in Table \ref{tab:events}, the corresponding observations could not be properly processed (i.e., the background brightness model could not be determined) due to the presence of extended bright objects in the FOV of the instruments (e.g., the Milky Way), saturated objects (e.g., a bright planet) and/or instrumental artifacts (e.g., ghost features). We kept only events with simultaneous observation in HI-1 both on-board STEREO-A and STEREO-B that allowed proper identification of the CME front in at least part of the FOV in each spacecraft. Due to these reasons, the following 6 events were removed from our analysis: $\#1$, $\#5$, $\#6$, $\#24$, $\#37$ and $\#38$. After removing these 6 events from the 20 available after the application of the criteria explained in Section \ref{sec:list}, we end up having 14 events.

\subsection{Extraction of the elongation profiles}
\label{sec:jmap}

To analyze the kinematic evolution of the events, we need to identify their corresponding fronts in the processed images 
and construct elongation-time maps of a given part of each front. The spatial location can then be derived under some assumptions for translating angular positions to heliocentric distances \cite{Sheeley2008a, Sheeley2008b,Rouillard2008,Rouillard2009,Rouillard2009b}.

\begin{figure}
 \noindent\includegraphics[width=\textwidth]{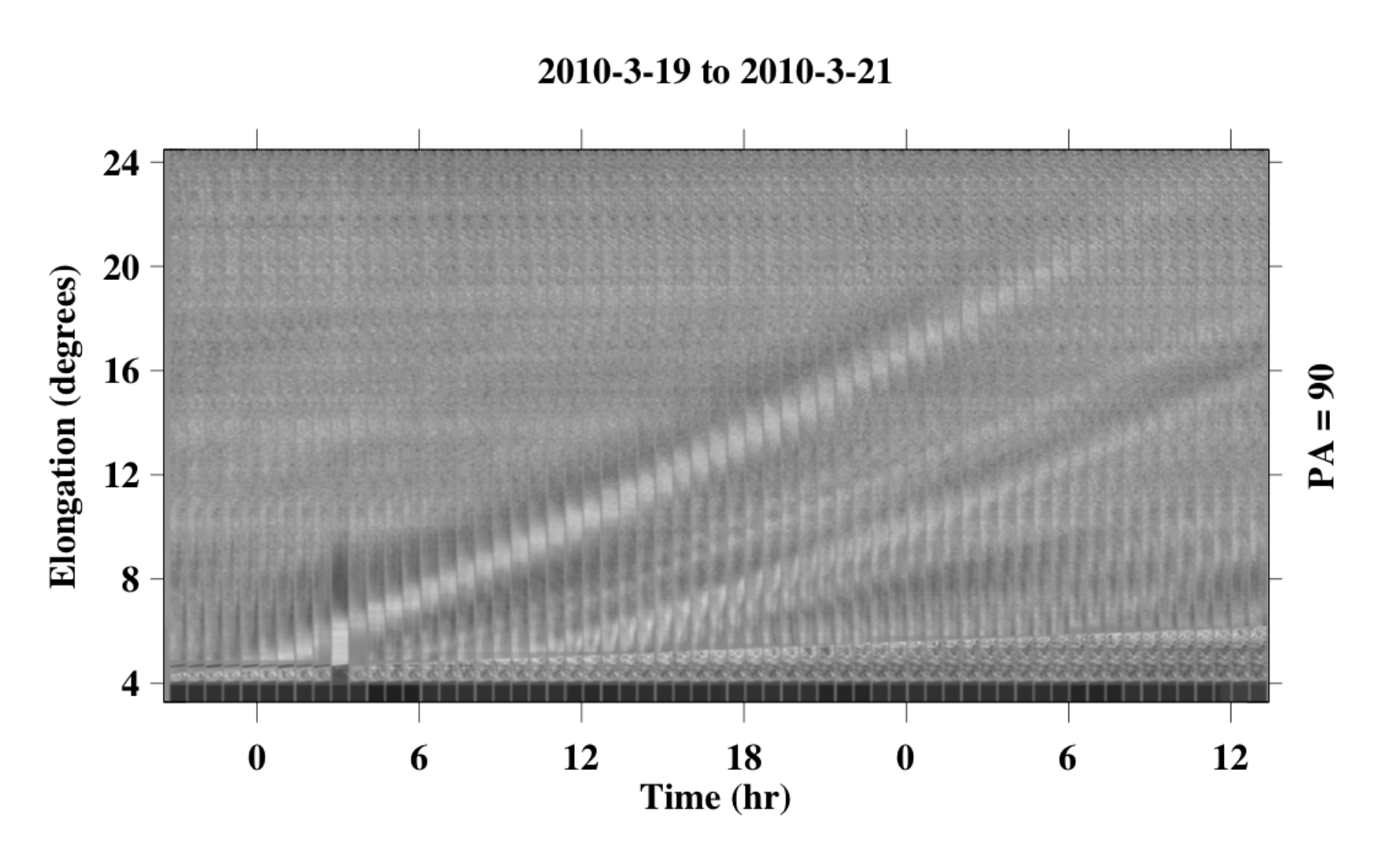}
\caption{Example of J-map of CME $\#1$ (2010/03/19) observed on HI-1-A FOV}
\label{fig:jmap_example}
\end{figure}

Given a set of sequential images observed by HI-1, we selected a position angle (PA) close to the ecliptic plane to construct the time-elongation profiles, frequently called J-maps \cite{Davies2009}. We use the PA of $90^{\circ}$ for STEREO-A and $270^{\circ}$ for STEREO-B, a region that nearly corresponds to the central height of the image. The PA is kept constant for a given CME event in each viewpoint, i.e., it is set to be the same at all instances.
Each time-elongation profile constructed in this way shows at least one bright feature that looks like an inclined line. This corresponds to the brighter points along the selected PA in the images, i.e., to the apex of the CME projected onto the plane of the sky at that particular PA. An example of a J-map created for event $\#1$ is shown in Figure \ref{fig:jmap_example}. Note that the brighter tracks in the map appear surrounded by a darker region. This is just a result of the computational processing applied to the images to reveal the CME features, which exploits the brightness contrast between the foreground and background in a way resembling an unsharp mask filter.

Once the J-map is created, we visually select the front. Since the identification is subjective, we repeat this procedure 9 times so that we can have an estimate of the error associated to the visual identification. In the following steps we normally take 3 time-elongation profiles: the median, minimum and maximum for each time instance (hereafter $\epsilon_{med}[t]$, $\epsilon_{min}[t]$, $\epsilon_{max}[t]$). 
A quick look at some events indicate that at the first time-instance $t_0$ we have $\epsilon_{max}[t_{0}]$ - $\epsilon_{min}[t_{0}] \approx 0.1^{\circ}$ and in the latest $t_f$ we get $\epsilon_{max}[t_{f}]$ - $\epsilon_{min}[t_{f}] \approx 0.3^{\circ}$. 
These 3 elongation versus time profiles are all used to estimate the CME Time-of-Arrival (ToA) at Earth, as described in Section \ref{sec:elliptical} and Section \ref{sec:results}.

In a few events, the J-maps produced at the PA mentioned ($90^{\circ}$ for STEREO-A and $270^{\circ}$ for STEREO-B) were not clear and we used PA shifted by up to 3 degrees instead. This happened due to the presence of  artifacts in the background at a given elongation, such as a bright planet. This negatively affected the CME front tracking in the J-map at that particular PA due to the excessive brightness of this feature as compared to that of both the background and the CME front. From our assessment using a few test CMEs, we understood that the shifted position angle within the range mentioned here produces differences that are within the error range between (from $\epsilon_{min}$ to $\epsilon_{max}$). Typically for a one degree PA shift, the elongation is changed by $0.1^{\circ}$ in the latest time-instance studied, generally lying in the range from $15^{\circ}-20^{\circ}$. Therefore, these shifts are not expected to affect significantly the results found here. 

\subsection{Overview of the CME Time-of-Arrival and Speed-on-Arrival determination}
\label{sec:model}

We calculate the travel time and Speed-on-Arrival of the CME using the drag model (Section \ref{sec:dragmodel}) and kinematic parameters derived from HI-1 observations from both spacecraft.

\begin{figure}
 \noindent\includegraphics[width=\textwidth]{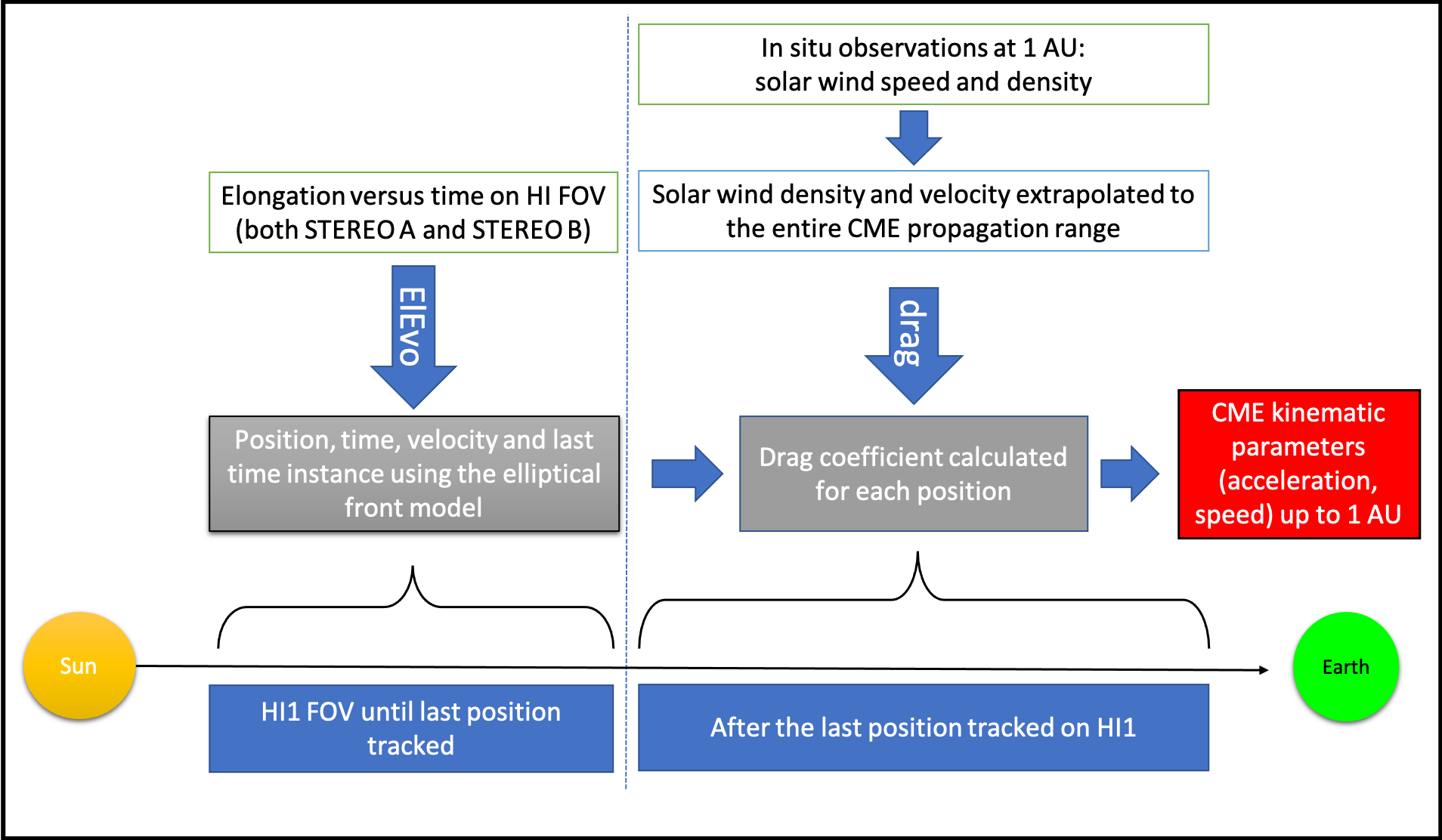}
\caption{Diagram explaining the CME's Time-of-Arrival (ToA) and Speed-on-Arrival (SoA) determinations. The left half illustrates the determination of CME front as a function of time from observation in the HI-1 FOV. The right half explains the application of the drag model, that is used only after the last position observed on HI-1 FOV. The boxes in white indicate inputs for the models and their outputs are shown in gray. The blue boxes indicate the range (along the Sun-Earth line) where each methodology is applied.}
\label{fig:diagram}
\end{figure}

The delineation of the procedure followed is depicted in the diagram in Figure~\ref{fig:diagram}. Briefly, we first extract the elongation of the CME front at a given PA as a function of time independently for each telescope. Then, a geometric model (Section \ref{sec:elliptical}) called Elliptical Conversion (ElCon) is used to derive the CME front position at each time instance, as well as its direction of propagation and its speed. These parameters are then used to calculate the CME acceleration at each point (in steps of $0.01$ au along the Sun-Earth line) after its last observation on HI-1 (typically from tenths of solar radii) to the L1 point (around $ 215\ R_\odot$) using the aerodynamic drag model (Section \ref{sec:dragmodel}). 

We have not considered other forces, such as the Lorentz force in this model, because this force is considered to be important only closer to the Sun, typically below $50\ R_{\odot}$ (the end of the HI-1 FOV is at about $96\ R_{\odot}$), especially for fast CMEs \cite{Bein2011, Sachdeva2015, Sachdeva2017}. 

 \subsection{The elliptical front model}
 \label{sec:elliptical}
 
\begin{figure}
\noindent\includegraphics[width=\textwidth]{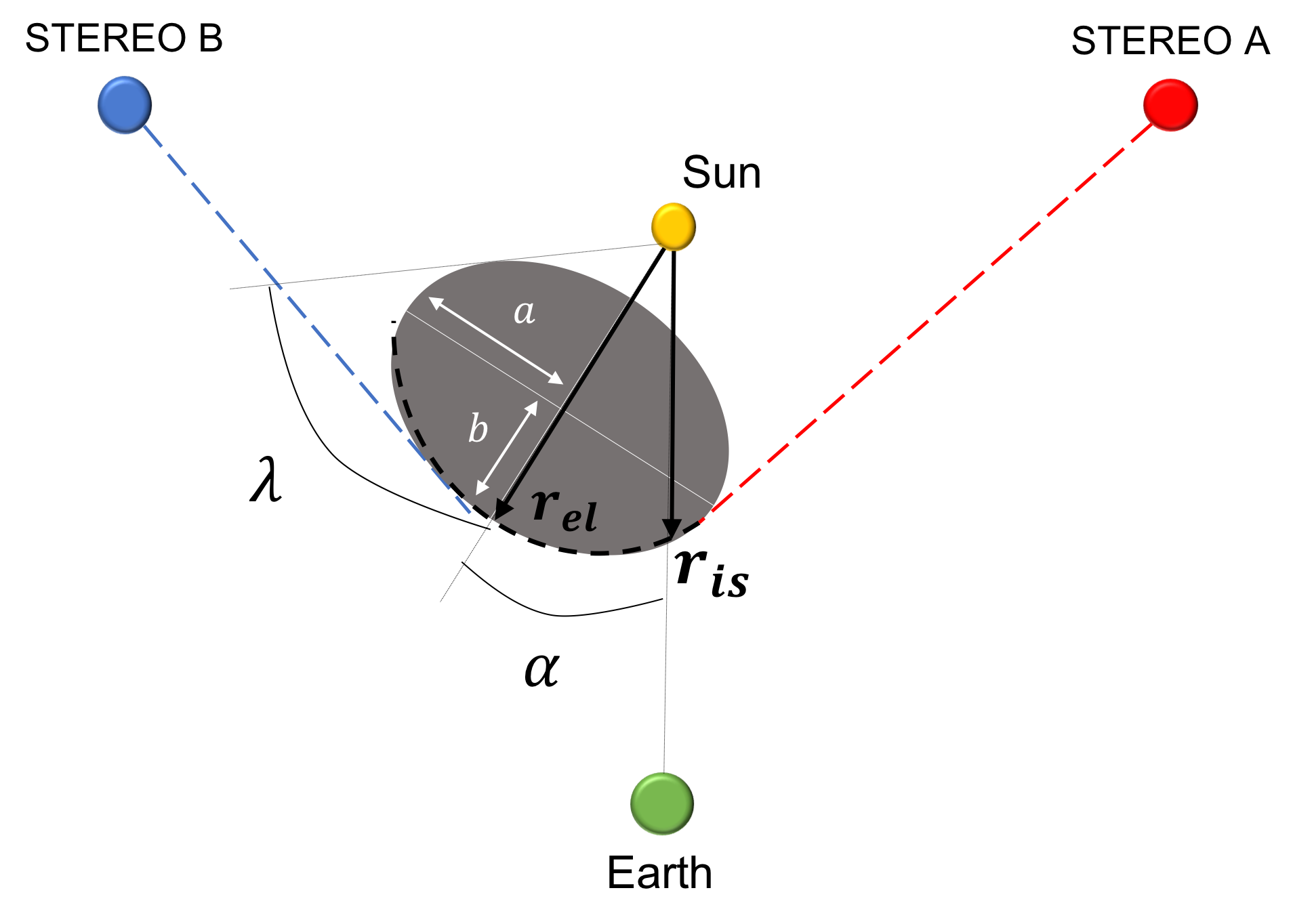}
\caption{The elliptical front model used in this study to derive the CME front position (dashed black line). The parameters of the model (CME half width in the ecliptic plane $\lambda$, aspect ratio of elliptical front $f=b/a$, central position angle in the ecliptic plane $\alpha$) are derived by a best-fit of a sequence of time-instances observed simultaneously by HI-1-A and HI-1-B. For each time-instance we derive the position of the CME front point along the Sun-Earth line ($r_{is}$) and along the central axis of the CME ($r_{el}$).}
\label{fig:elliptical}
\end{figure}

 To derive the CME position in the HI-1 FOV, we adopt the Elliptical Conversion (ElCon) model as described in \citeA{Mostl2015} and \citeA{Rollett2016}. This model considers an elliptically shaped CME front on the ecliptic plane. Its position and speed can then be derived at any location in space using just geometrical arguments, provided the time evolution of the front's elongation is known and a set of given parameters of the CME front (e.g., angular width, direction of propagation, aspect ratio, etc.) are defined. The model adds an extra degree of freedom when compared to circular CME fronts, which is the aspect ratio. Since the CMEs can have various shapes, the elliptical front is a more general fit allowing more CMEs to be fit. This model and its parameters are shown in Figure \ref{fig:elliptical}.
 
 In this study, we derive the parameters of the model (CME half width in the ecliptic plane $\lambda$, aspect ratio of elliptical front $f$, central position angle in the ecliptic plane $\alpha$, speed) by doing a best-fit of a sequence of $n$ time-instances observed simultaneously by HI-1-A and HI-1-B. In previous studies these parameters were fixed for a given set of CMEs for simplicity \cite{Mostl2015,Rollett2016}. The residual $\sigma$ was calculated from the following expression:
 
  \begin{linenomath*}
 $\sigma= \sum_{t=1}^{n}(| r_{is}(t)^{A}-r_{is}(t)^{B}| + | r_{el}(t)^{A}-r_{el}(t)^{B}|)/n$

  \end{linenomath*}
 
 where $r_{is}(t)$ is the position of the CME front point along the Sun-Earth line on the ecliptic plane as a function of time, $t$, derived using observations from a given spacecraft; $r_{el}(t)$ denotes the position of the central axis on the CME front. The angle between $r_{is}$ and $r_{el}$ corresponds to $\alpha$. Positive values of $\alpha$ indicate that the CME propagates towards STEREO-A.  The superscripts $A$ and $B$ indicate values from STEREO-A and STEREO-B observations, respectively.  

The list of parameters derived using the ElCon model is shown in Table \ref{tab:elevoparams}. As already mentioned, from the list of 38 events shown in Table \ref{tab:events}, only 14 are used with the ElCon model. The rest were removed due to the reasons described in Section \ref{sec:list} and~\ref{sec:removal}.

   \begin{table}
  \caption{Parameters of the CMEs elliptical front derived using the ElCon model: linear speeds ($v_{med}$, $v_{min}$ and  $v_{max}$), CME half width in the ecliptic plane ($\lambda$), aspect ratio of the front ($f$) and the CME central position angle on the ecliptic plane ($\alpha$, positive ahead of the Earth). Other parameters shown are the residual ($\sigma$) and the position of the last point that the CME was tracked simultaneously on both viewpoints ($s_0$).}
 \label{tab:elevoparams}
\begin{tabular}{l c c c c c c c c c c}
 \hline
  ID& last tracked time (UT) & $s_0$ & $v_{med}$ & $v_{min}$ & $v_{max}$ & $f$ & $\lambda$ & $\sigma$ & $\alpha $  \\
  & (UT) & [au] & [km/s] &  [km/s] & [km/s] &  & $[^{\circ}]$ &  [au] & $ [^{\circ}]$  \\
 \hline
$   2$&  03-Apr-2010 20:29:21 &  0.22  &  846  &  866  &  876    & 0.5   &  80 &0.0047  &  -17 \\
$   3$&  09-Apr-2010 00:39:22 &  0.28  &  490  &  448  &  491    & 0.6   &  60 &0.0021  &   12 \\
$   9$&  15-Feb-2011 18:29:34 &  0.26  &  465  &  456  &  475    & 0.5   &  65 &0.0028  &  -12 \\
$ 11$& 26-Mar-2011 07:59:25 &  0.20  &  448  &  446  &  427    & 0.5   &  50 &0.0033  &  -11 \\
$ 13$& 14-Jun-2011 23:49:28 &  0.28  &  769  &  765  &  775    & 0.5   &  70 &0.0081  &  -19 \\
$ 17$& 14-Sep-2011 10:29:53 &  0.14  &  605  &  584  &  568    & 0.5   &  80 &0.0093  &   63 \\
$ 25$& 20-Apr-2012 10:29:25 &  0.25  &  446  &  446  &  453    & 0.5   &  80 &0.0079  &  -32 \\
$ 26$& 15-Jun-2012 03:19:22 &  0.27  &  741  &  755  &  776    & 0.5   &  80 &0.0051  &   -7 \\
$ 27$& 13-Jul-2012 07:59:27 &  0.31  &  743  &  732  &  780    & 0.9   &  80 &0.0475  &   20 \\
$ 28$& 28-Sep-2012 08:29:50 &  0.18  &  740  &  721  &  739    & 0.6   &  20 &0.0056  &   20 \\
$ 29$& 06-Oct-2012 01:49:52 &  0.36  &  692  &  686  &  712    & 0.5   &  30 &0.0075  &   15 \\
$ 30$& 28-Oct-2012 11:59:57 &  0.21  &  431  &  422  &  441    & 0.5   &  20 &0.0097  &   11 \\
$ 33$& 15-Mar-2013 15:59:43 &  0.19  &  765  &  703  &  737    & 0.6   &  80 &0.0030  &    1 \\
$ 34$& 11-Apr-2013 15:49:33 &  0.18  &  764  &  780  &  667    & 0.5   &  65 &0.0042  &   -1 \\
 \hline
    \end{tabular}
  \end{table}

We fitted the elliptical model three times for each event: one using the median elongation extracted at each time instance $\epsilon_{med}[t]$, a second with minimum elongation $\epsilon_{min}[t]$ and a third time using the maximum $\epsilon_{max}[t]$. In each case, a set of parameters $\lambda$, $f$, $\alpha$ is derived and $r_{is}$ at each time instance $t$ is calculated as the average of $r_{is}^{A}$ and $r_{is}^{B}$. 

$v_{med}$, $v_{min}$, $v_{max}$ are the linear speeds calculated from the parameters of the ElCon derived using $\epsilon_{med}[t]$, $\epsilon_{min}[t]$ and $\epsilon_{max}[t]$, respectively. All three speeds considered here are calculated along the Sun-Earth line, i.e., using $r_{is}$. The differences between the 3 values (typically well below $ 50\ km/s$) give us an idea of the error introduced in the CME speed due to differences in the identification of the CME front in the J-maps. The 3 speeds are used for the calculation of the CME travel time and ToA error, as described in Section \ref{sec:results}. In Section \ref{sec:comparing_speeds}, we compare speeds derived in this work with previous studies. The remaining parameters shown in Table \ref{tab:elevoparams} ($f$, $\lambda$, $\sigma$ and $\alpha$) are calculated using the median elongation profile.

The elliptical fronts we derived correspond to wide CMEs in most cases: 11 of the 14 events have $\lambda \geq 50^{\circ}$. The CMEs aspect ratio ranges from $0.5$ to $0.6$ for all cases, except for \#27, which is $0.9$. This means that  all CME fronts are elongated perpendicular to the propagation direction. 

There is limited literature to compare the elliptical front geometries derived here \cite{Rollett2016, Mostl2015}. In \citeA{Rollett2016} both $f$ and $\lambda$ were set to fixed values for all CMEs in their set of 21 events to test their methodology. The aspect ratios ($f$) derived here are typically lower than values assumed on \citeA{Rollett2016}, which are $f=0.8$, $f=1.0$ and $f=1.2$. It is interesting to note that their results suggest that among the 3 values of $f$ used, $f=0.8$ leads to smaller SoA errors, which is the closest to the average aspect ratio found in our study ($f=0.55$). The SoA errors found in \citeA{Rollett2016} are: $17\pm54\ km/s$ ($f=0.8$), $21\pm63\ km/s$ ($f=1$) and $38\pm87\ km/s$ ($f=1.2$). Regarding ToA, \citeA{Rollett2016} found the smallest error with $f=1.2$ ($5.0\pm5.6\ h$). Using $f=0.8$, the error is approximately 1 hour higher ($6.3\pm5.5\ h$). Since the objective of \citeA{Rollett2016} work was to introduce the methodology and assess its reliability, they set the half-width of all events under study to $35^{\circ}$ to simplify their analysis. Thus, it is impossible to derive conclusions by comparing our results with theirs. \citeA{Mostl2015} considered a single CME event observed on January 7, 2014, which is not in our list. Due to specific reasons associated to this event, including the in situ observations that provide some constraints on the CME geometry, the half-width calculated is in the range from $35^{\circ}$ to $60^{\circ}$ and the front ratio $f$ ranges from $0.55$ to $1.0$. The results in \citeA{Mostl2015} are in the same range as ours. Overall, the elliptical front geometries we derived are consistent with previous studies.

 \subsection{The drag model}
\label{sec:dragmodel}

The aerodynamic drag results from the interaction of the CME with the solar wind. There are many works that apply such kind of force, most of them relying on empirically-derived drag coefficients \cite{Cargill2004,Vrsnak2006,Howard2007,Borgazzi2009,Byrne2010,Maloney2010,Vrsnak2010,Vrsnak2013,Mishra2013,Dolei2014,Iju2014,Temmer2015}. 

Among these works, many authors have used a constant drag coefficient for a given CME in its path from the Sun to the Earth so that the drag could only be a function of the (i) difference between the CME and solar wind speed, (ii) CME mass, (iii) solar wind density, and (iv) cross section area of the CME \cite{Vrsnak2013, Temmer2015, Mishra2013}. Only a few works have used a drag coefficient as a function of the Reynolds number, which in turn depends on the viscosity of the solar wind plasma \cite{Subramanian2012,Sachdeva2015}. 

The drag force description using the Reynolds number was found to work quite well for CMEs analyzed in \citeA{Subramanian2012}. We believe that this method based on a physical description of the plasma is a better solution than using either ad hoc or empirical parameters, which are normally derived using a set of CMEs. As background solar wind conditions are dramatically different from case to case, some events may not have their particularities represented in the set of events used to define the empirical parameters and, therefore, they may not be appropriately described.

Following \citeA{Sachdeva2015}, we consider the drag force description given by:

\begin{linenomath*}
$F_{drag} [s]=-m_{CME}\   \gamma [s]\  (v_{CME} [s]-v_{SW} [s])\ |v_{CME} [s]-v_{SW} [s]|,$
\end{linenomath*}

where $v_{CME}$ is the CME speed and $v_{SW}$ is the background solar wind speed and $\gamma$ is the drag parameter. Both speeds are a function of CME position $s$ along the Sun-Earth line. For a CME propagating  towards the Earth, $s$ increases as time passes.  $m_{CME}$ is the CME mass taken from the CDAW CME catalog \cite{Yashiro2004}. If not available, we consider $m_{CME}=1.1\times{10}^{15} g$, the median value reported on \citeA{Vourlidas2010} for CMEs observed between 1996 and 2009.

\protect The drag force adopted here, which is proportional to the square of difference between the CME speed and the solar wind speed, is also used in several studies involving observations from STEREO \cite<see, e.g.,>{Sachdeva2015, Sachdeva2017, Maloney2010, Mishra2013, Temmer2012, Mishra2014, Salman2020, Hess2015}.

The drag force can also be described as proportional to $(v_{CME}-v_{SW})^\beta$ with $\beta$ set to one, two or determined empirically. Considering the ToA, \citeA{Vrsnak2002} found smaller errors with $\beta=1$ for events observed by LASCO while \citeA{Shanmugaraju2014} found the smaller errors with $\beta=2$ for a set of CMEs observed by STEREO. \citeA{Byrne2010} empirically found $\beta=2.27$. \citeA{Shi2015} found that $\beta=2$ results in a better CME ToA prediction than the linear one for a set of 21 CMEs observed by STEREO. \citeA{Shi2015} also considered a hybrid model that combines both $\beta=1$ and $\beta=2$ descriptions and found that the latter has a larger contribution in the ToA determination. 

Here we adopt the description of the drag based on a physical model of the viscosity mechanism, as was done by \citeA{Subramanian2012}, \citeA{Sachdeva2015} and \citeA{Sachdeva2017}. The drag that is proportional to $v_{CME}-v_{SW}$ is normally used in studies that are focused on empirical descriptions of the dynamics of the CME based on the observed CME speed profiles. Our objective here is not following an empirical description of the drag model and, for this reason, we adopt the description of the drag force proportional to the square of the CME speed.

Here we consider that $\gamma$ is given by:

\begin{linenomath*}
$\gamma\left[s\right]=C_D\left[s\right]  n_{SW}\left[s\right]\frac{m_{P} A_{CME}\left[s\right]}{m_{CME}}$	
\end{linenomath*}

where $C_D$ is the dimensionless drag coefficient, $n_{SW}$ is the solar wind proton number density, $m_P$ is the proton mass, $A_{CME}$ is the CME cross section area (explained in the next paragraphs) and $m_{CME}$ is the CME mass. Typically $\gamma$ has values ranging from $1 \times 10^{-9}\ km^{-1}$ to $2 \times 10^{-7}\ km^{-1}$, see, e.g. \citeA{Temmer2015} and \citeA{Vrsnak2013}.

In several previous studies, $C_D$ was empirically determined and considered to be constant (see, e.g. \citeA{Cargill2004}; \citeA{Vrsnak2010}; \citeA{Mishra2013}; \citeA{Temmer2015}, and references therein). In these studies, $C_D$ typically ranges from $0.2$ to $0.4$. 

In this study, on the other hand, we determine the value of $C_D$ using a set of equations based on a physical definition of the CME aerodynamic drag introduced by \citeA{Subramanian2012} and previously studied by \citeA{Sachdeva2015, Sachdeva2017}. Here we describe $C_D$ using the following expression determined experimentally by \citeA{Achenbach1972}:

\begin{linenomath*}
$C_D\left[s\right]=0.148-4.3\times{10}^4 (R{e\left[s\right]})^{-1}+9.8\times{10}^{-9} Re[s]	$	.
\end{linenomath*}

This equation for $C_D$ is a fit to data observed on a solid metal sphere immersed in a flow with high Reynolds number $Re$. We considered that this result is suitable for the interaction of the CME with the background solar wind because (i) the equation of the drag force considers a solid-like body immersed on a high-Reynolds number and (ii) typically the boundaries of magnetic clouds (and therefore, CMEs) are over-pressured structures, i.e., they have a substantial jump in their total pressure (magnetic plus plasma) in the region close to their boundaries \cite{Jian2006}.

The Reynolds number depends on the macroscopic lengthscale of the CME, its velocity relative to the background solar wind particles and the viscosity of the solar wind. For more details, the reader is referred to \citeA{Sachdeva2017}.

The  CME cross section area $A_{CME}$ is calculated as:

\begin{linenomath*}
$A_{CME}\left[s\right] = \pi \times \ R_{CME}^2\left[s\right] \times w / 360 $	
\end{linenomath*}

where $w$ is the width of the CME (in degrees, as determined by the CME CDAW catalog) and $R_{CME}$ is the radius of the CME that was taken to be $0.4  s$. This expression of $R_{CME}$ was experimentally chosen in this study as a good solution to reduce the ToA error for the set of CME events studied here among different values of the coefficient lower than the unit. 

\subsection{Background solar wind speed}
\label{sec:background_speed}

As described in previous section, the solar wind speed $v_{SW}$ at any point along the Sun-Earth is required to calculate the drag. Close to 1 au, the solar wind conditions are continuously observed by instrumentation at the Lagrangian point L1, such as by the  Solar Wind Electron, Proton, and Alpha Monitor (SWEPAM) instrument \cite{McComas1998} onboard Advanced Composition Explorer (ACE) mission \cite{Stone1998} and by the Solar Wind Electron (SWE) instrument \cite{Ogilvie1995} onboard Wind spacecraft \cite{King2005}. In the remaining points of the trajectory, on the other hand, $v_{SW}$ needs to be calculated using empirical models or simulation.

In this study, we use an empirical expression to extrapolate the solar wind speed at any position along the Sun-Earth line using observation at 1 au ($v_{SW@1au}$). Following \citeA{Sheeley1997} and \citeA{Sheeley1999}, the solar wind speed along the Sun-Earth line ($v_{sw}[s]$) is considered to be:

\begin{linenomath*}
$v_{SW}^2[s]=v_{SW@1au}^2[1-e^{-(s-r_{0})/r_{a}}]$
\end{linenomath*}
	
where $s$ is a given position along the Sun-Earth line, $r_{0}=1.5 \ R_{\odot}$ is the distance from the Sun where the solar wind is taken to be zero and $r_{a}=50\  R_{\odot}$ is the distance over which the asymptotic speed is reached. According to this model, the solar wind speed increases more significantly close to the Sun, typically  up to approximately  $100\ R_{\odot}$, and then it is almost constant up to $1\ au$. 

In this work, we considered that $v_{SW@1au}$ is the average observed value in the time period from 48 up to 24 hours before the CME is first observed on the LASCO/C2 FOV. We chose this time period taking into account the typical travel time for a solar wind parcel to travel from the solar corona to 1 au.

\subsection{Background solar wind density}
\label{sec:background_wind}

Besides the solar wind speed, the solar wind density along the CME trajectory is also required for calculating the drag force as described in Section \ref{sec:dragmodel}. Again, the observations are limited to 1 au and the density evolution must be derived via a model.  The solar wind proton density $n_{SW}$ as a function of position $s$ is given by \citeA{Leblanc1998}:

\begin{linenomath*}
$n_{SW}\left[s\right]=\left(\frac{n_{SW@1au}}{7.2}\right)\left(3.3\times{10}^5 s^{-2}+4.1\times{10}^6{  s}^{-4}+8\times{10}^7{  s}^{-6}\right)$ 
\end{linenomath*}	

where $n_{SW@1au}$ is the solar wind density observed in the L1 Lagrangian Point (close to 1 au). Here we use the model of electron density from \citeA{Leblanc1998} assuming that the electron and proton densities are equal. The term between parentheses considers the difference of the density at $1\ au$ from the original value of $7.2\ cm^{-3}$ used on the model. $n_{SW@1au}$ was assumed to be the average observation value from 48 up to 24 hours before the CME first observation on LASCO C2.

In the density equation, $s^{-2}$ is the dominant term in the outer corona and inner heliosphere. We, nevertheless, retain the full expression for completeness.

 \section{Results} 
 \label{sec:results}

\protect Now we explain and exemplify the application of the drag force to derive the CME speeds and profiles as a function of position (Section \ref{sec:results_app_drag}). Then, we compare the speeds we derived with previous studies that include some CMEs studied here (Section \ref{sec:comparing_speeds}) and one catalog based on observations from HI-1 (Section \ref{sec:comparing_with_helcats}). Finally, we show results from the ToA and SoA errors in Section \ref{sec:results_toa_erros} and \ref{sec:Soa_and_error}.

\subsection{Application of the drag model}
\label{sec:results_app_drag}

To calculate the drag coefficient, we use the last HI-1 observation position for which the CME front is visible and the linear speed of the portion of the CME front along the Sun-Earth line ($v_{med}$, $v_{min}$, $v_{max}$). Some geometric parameters derived using ElCon (such as the angular width and angle) are not used explicitly in the drag force model, but they are indirectly taken into account in the derivation of $r_{is}$ at each time-instance. 

We start the application of the drag model at the last HI-1 observation position ($s_0$). In some cases, the brightness of the CME front is similar to that of the background (specially in the outer half of the FOV).  In these cases, the CME front cannot be resolved. Thus, the $s_0$ position changes from event to event ($s_0$ is indicated in the second column of Table \ref{tab:elevoparams}). Typically, the last height-time observation ranges between 20 and 80 solar radii.

\begin{figure}
 \centering
 \noindent\includegraphics[width=\textwidth,height=15cm,keepaspectratio]{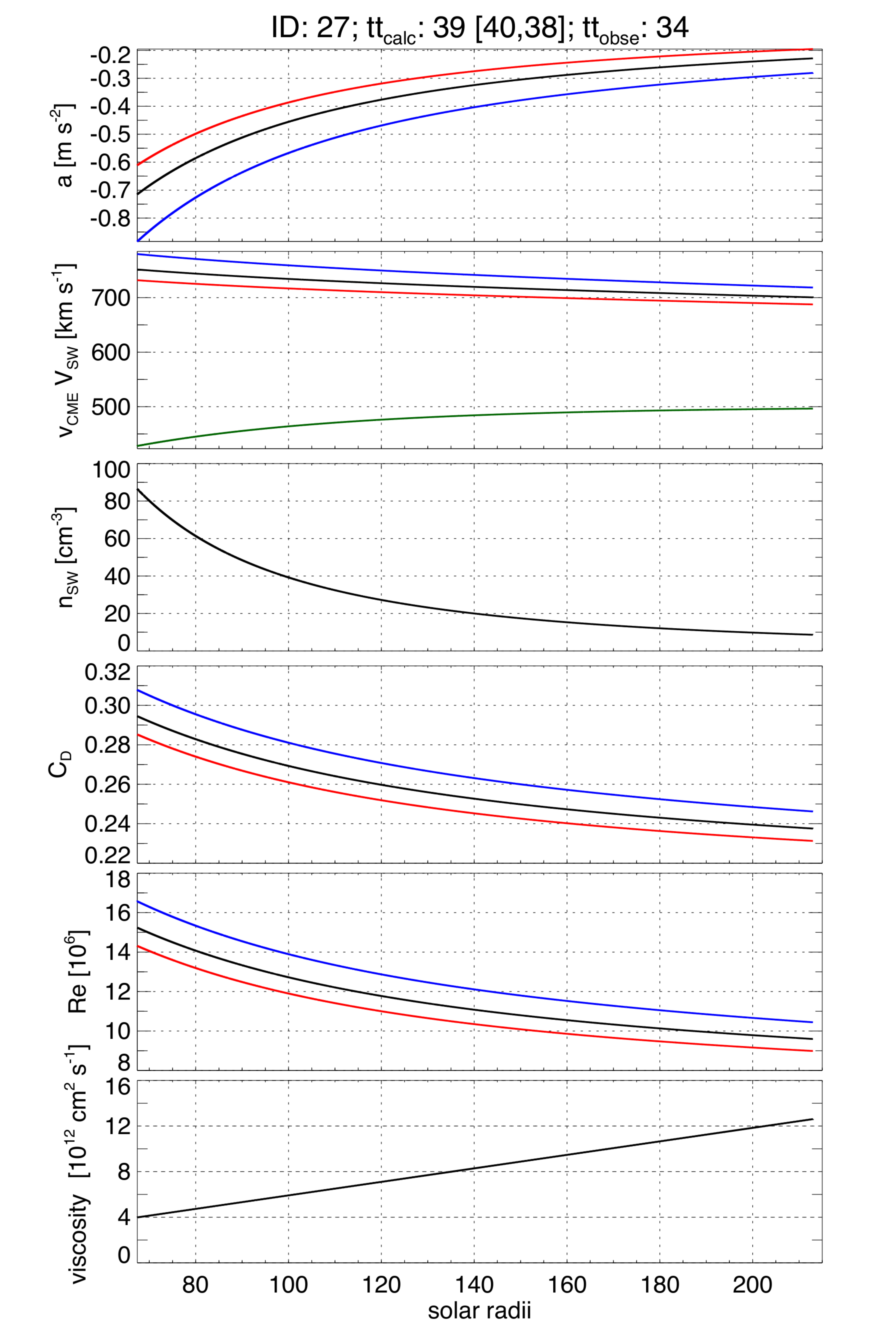}
\caption{The application of the drag force to a sample CME that is decelerated from the solar corona to 1 au. In the panels with multiple lines, the black ones indicate the CME kinematic parameters calculated using $v_{med}$ and the red and blue lines indicate CME parameters calculated using $v_{min}$ and $v_{max}$, respectively. The green line on the second panel (from top to bottom) indicates the background solar wind speed.}
\label{fig:dragexa}
\end{figure}

An example of the application of the drag model is shown in Figure \ref{fig:dragexa}. In each panel, the horizontal axis shows the distance from the Sun (in solar radii). In the top left panel, the black line denotes the acceleration based on the initial speed $v_{med}$. Acceleration profiles based on $v_{min}$ and $v_{max}$ are indicated by the red and blue lines, respectively. The CME speed derived using $v_{med}$ as initial speed is shown in the second panel, from top to bottom. The speed of the background solar wind speed ($v_{sw}$) is indicated by the green line. The remaining lines represent speeds calculated using $v_{min}$ and $v_{max}$. The background solar wind proton density is represented on the third panel, from top to bottom. Other parameters shown are the drag coefficient (fourth panel, from top to bottom), the Reynolds number (fifth panel) and the viscosity (lower panel).

For all analyzed events, $C_D$ has a decreasing profile from the Sun to 1 au, typically with steeper slope close to the Sun, as shown in Figure \ref{fig:dragexa}. The variations for the different cases arise from differences between the CME and background solar wind speed and density, and the CME area and mass. Close to the Sun, $C_D$ ranges from $0.36$ to $0.19$ while at L1 its values ranges from $0.16$ to $0.28$. Values of $C_D$ in any position mentioned above lie in the same range than previous studies that adopted a single drag coefficient for a set of events, which have values typically chosen between $0.2$ and $0.4$.

As the CME moves toward the Earth, the background solar wind speed increases asymptotically. Given the nature of the drag force, the CME decelerates and, as a result, the magnitude of the drag force decreases.  Other reasons for the decrease in the drag force with distance are: (i) the solar wind density ($n_{sw}$) decreases (from values typically around $50\ cm^{-3}$ to $5\ cm^{-3}$ and/or (ii) the Reynolds number ($Re$) decreases thus reducing the drag coefficient $C_{D}$.

All 14 CMEs in our sample decelerate since all have $v_{CME}> v_{SW}$. The deceleration  rate is higher close to the Sun (values up to $3.25\  m/s^{2}$) and decreases as the CME propagates toward $1\ au$.

\subsection{Comparison of the CME speeds with previous studies}
\label{sec:comparing_speeds}

We now compare the speeds we derived here with past works. Several CMEs in our list were analyzed elsewhere \cite{Mostl2014,Rollett2016,Barnard2017, Colaninno2013, Wood2017}. However, most of these studies considered coronagraph observations alone or coronagraph observations combined with heliospheric images. Thus, the speed derived by them is typically at an earlier stage of the CME propagation than done here. Only \citeA{Rollett2016} used observations exclusively from heliospheric imagers.

Table \ref{tab:comparing_speeds} compares the speed measurements across the various studies. The first column refers to the event number in this manuscript (Tables \ref{tab:events} and \ref{tab:elevoparams}). The second column lists the reference and event ID in that reference. The references for each event vary since each study used its own criteria for CME selection. The speeds are listed in the third column and the fourth column contains some remarks about the particularities of the speed derived in each case.

\citeA{Mostl2014} has 8 CMEs in common with our study. They use the average speed between $2.5$ and $15.6$ $R_{\circ}$ based on  Graduated Cylindrical Shell (GCS) model fits \cite{Thernisien2006, Thernisien2009, Thernisien2011}. The derived speeds are not necessarily on the ecliptic plane, which is where our speeds are derived. Another difference is that we track the CMEs in the HI-1 FOV while \citeA{Mostl2014} considers only the coronagraph FOV. We see that fast CMEs have higher speeds in \citeA{Mostl2014}. This is not surprising since their speeds are derived lower in the corona and CMEs tend to decelerate away from the Sun \protect \cite<see the review from>[and references therein]{Manchester2017}. So, the speed differences between the two works likely reflect deceleration rather than any measurement discrepancies.

\citeA{Sachdeva2017} also derived the CME kinematics with the GCS fitting technique but using observations from both the coronagraphs (including LASCO) and heliospheric imagers. Although we are using the same events, we derive the CME initial speed from a third-degree polynomial fit to the height-time observations of the CME leading edge. Another difference is that the CME speed derived on \citeA{Sachdeva2017} is not necessarily on the ecliptic plane; in some cases the GCS model can be more than $30^\circ$ away in Carrington latitude. We identified a significant difference in the speeds for events \#26, \#27, \#28 and \#34 and a reasonable agreement for the remaining 10 events. We consider \say{reasonable} differences of the order of $100\ km/s$ or less, considering the typical difference between CME speed derived by different methods \cite<see, e.g., >{Mierla2010}. In the other 4 cases, our speeds are significantly lower than those obtained by \protect \citeA{Sachdeva2017}. The difference may be due to the height where those speeds refer to. \citeA{Sachdeva2017} derives the initial speed at less $10$ $R_{\circ}$ (see remarks in the third column of Table \ref{tab:comparing_speeds}) while our speeds are the average speeds  in the HI-1 FOV only, typically up to $40-60$ $R_{\circ}$. CMEs with speeds exceeding $1000\ km/s$ (as is the case for events \#26, \#27, \#28 and \#34) are expected to decelerate compared to their speed at 10 solar radii \cite{Sachdeva2017}.

\citeA{Rollett2016} applies the same ElCon method as us in the HI-1 FOV  but using only single viewpoint observations from STEREO-A.  In contrast to our study, \citeA{Rollett2016} use the fixed-$\phi$ method to derive some parameters of the ElCon model, such as direction of propagation. Another difference is that their speeds are derived at positions closer to the Sun than ours. Many speeds from \citeA{Rollett2016} mentioned in Table \ref{tab:comparing_speeds} are derived doing a fit of the drag-based model to the initial point in the trajectory of the CME studied. In some events, this point is below $10$ $R_{\circ}$. This difference can partially explain why the speeds derived by \citeA{Rollett2016} are higher than ours for the fastest CMEs.

\citeA{Barnard2017} estimated the speed of CME \#29 in the coronagraphs FOV using the CME analysis tool (CAT) \cite{Millward2013}, which assumes a \say{teardrop}-shaped CME (similar to a cone model but with curved leading edge). Similarly to the previous references, the CAT speed is derived much closer to the Sun than ours (at $77$ $R_{\circ}$ in this particular case). However, the difference in the two speeds is only $ 6\ km/s$. 

\citeA{Gopalswamy2013} measured the speeds on the STEREO/COR2 coronagraph (i.e., up to about $15$ $R_{\odot}$) in the ecliptic plane. Therefore, speeds from this reference are located much closer to the Sun than our speeds. \citeA{Gopalswamy2013} consider their speeds to be unprojected because they were measured when the STEREO spacecraft position were within $30^\circ$  from quadrature. In each event, they selected the STEREO spacecraft (Table \ref{tab:comparing_speeds}) with CME observations closer to the limb. We have 4 common CMEs (\#9, \#17, \#26 and \#27). For all four CMEs, our derived speeds are significantly smaller (by hundreds of $km/s$) than in \citeA{Gopalswamy2013}. We believe that the same explanation holds for these discrepancies; namely, the faster CME decelerate as they travel away from the Sun.

To summarize, the speed comparisons with our method suggest that the height where speeds are measured plays a very big role, particularly for faster CMEs. Our speeds are lower than the works in Table~3 because we measure the CME kinematics at a later stage in their propagation, when they have undergone deceleration.

\begin{table}
    \centering
  \caption{Comparison of speeds derived for CMEs studied here with previous studies.}
 \label{tab:comparing_speeds}
 \scalebox{0.8}{

\begin{tabular}{c c c c c }
 \hline
CME \# &	Reference and event number & Speed  & Remarks\\
 \hline
$2$     &	this study	                & $846\ km/s$   &up to $47$ $R_{\circ}$\\
        &	\citeA{Mostl2014}, \# 7	    &$829\ km/s$    & initial speed   \\
        &	\citeA{Rollett2016}, \# 5	&$1145\ km/s$   & $35.8$ $R_{\circ}$, $f=1$	\\
        &	\citeA{Wood2017}, \#8       &$960\ km/s$    & peak speed  	     \\
        &	\citeA{Wood2017}, \#8       & $660\ km/s$   & terminal speed 	 \\
        &   \citeA{Sachdeva2017}, \#2	& $916\ km/s$   & $5.5$ $R_{\circ}$ \\
\hline
$3$	    &   this study 	                & $490\ km/s$   & up to $60$ $R_{\circ}$\\
        &   \citeA{Mostl2014}, \#8	    & $511\ km/s$   & initial speed	\\
        &   \citeA{Rollett2016}, \#6    &	$989\ km/s$ &$3.6$ $R_{\circ}$ $f=1$ &	 \\
    	&   \citeA{Sachdeva2017}, \#3	& $506\  km/s$  &$19.7$ $R_{\circ}$&	\\
\hline
$9$ 	&   this study	                & $465\ km/s$   & up to $56$ $R_{\circ}$\\
     	&   \citeA{Mostl2014}, \#13	    & $557\ km/s$   & initial speed\\
     	&   \citeA{Rollett2016}, \#10   & $720\ km/s$   & $20.7$ $R_{\circ}$ $f=1$\\
     	&   \citeA{Wood2017}, \#15	    & $1388\ km/s$  &peak speed\\
     	&   \citeA{Wood2017}, \#15      & $557\ km/s$   &terminal speed\\
     	&   \citeA{Gopalswamy2013}, \#4 & $864\ km/s$   &COR2, STEREO A\\
     	&   \citeA{Sachdeva2017}, \#9	& $530\ km/s$   &$39.7$ $R_{\circ}$\\
\hline
$11$	&this study                     &$448\ km/s$        & up to $43$ $R_{\circ}$\\
    	&\citeA{Wood2017}, \#16	        &$352\ km/s$        &peak speed\\
    	&\citeA{Wood2017}, \#16         &$352\ km/s$        &terminal speed\\
     	&\citeA{Sachdeva2017}, \#11	    &$456\ km/s$        &$46.5$ $R_{\circ}$\\
\hline
$13$	&this study	                    &$769\ km/s$    &up to $60$ $R_{\circ}$\\
    	&\citeA{Wood2017}, \#19	        &$789\ km/s$    &peak speed\\
    	&\citeA{Wood2017}, \#19         &$789\ km/s$    &terminal speed\\
     	&\citeA{Sachdeva2017}, \#13 	&$767\ km/s$    &$24.4$ $R_{\circ}$\\
\hline
$17$	&this study                     &$605\ km/s$    &up to $30$ $R_{\circ}$\\
    	&\citeA{Wood2017}, \#21         &$518\ km/s$    &peak speed\\
     	&\citeA{Wood2017}, \#21         &$518\ km/s$    &terminal speed\\
     	&\citeA{Gopalswamy2013}, \#11   &$467\ km/s$    &COR2, STEREO B\\
     	&\citeA{Sachdeva2017}, \#17	    &$636\ km/s$    &$38.8$ $R_{\circ}$\\
\hline
$25$	&this study	                    &$446\ km/s$   &up to $54$ $R_{\circ}$\\
    	&\citeA{Mostl2014}, \#22	    &$639\ km/s$   &initial speed\\
     	&\citeA{Rollett2016}, \#19	    &$625\ km/s$   &$18.8$ $R_{\circ}$ $f=1$\\
     	&\citeA{Sachdeva2017}, \#25	    &$684\ km/s$   &initial speed at $23.1$ $R_{\circ}$ \\
\hline
$26$	&this study                     &$741\ km/s$    &up to $58$ $R_{\circ}$\\
    	&\citeA{Mostl2014}, \#23	    &$1102\ km/s$   &initial speed\\
     	&\citeA{Rollett2016}, \#20      &$1438\ km/s$   &$15.6$ $R_{\circ}$ $f=1$\\
     	&\citeA{Wood2017}, \#26         &$1104\ km/s$   &peak speed\\
     	&\citeA{Wood2017}, \#26         &$658\ km/s$    &terminal speed\\
     	&\citeA{Gopalswamy2013}, \#19   &$1317\ km/s$   &COR2, STEREO B\\
     	&\citeA{Sachdeva2017} , \#26	&$1152\ km/s$   &initial speed at $6.2$ $R_{\circ}$\\
\hline
$27$	&this study                     &$743\ km/s$    &up to $67$ $R_{\circ}$\\
    	&\citeA{Mostl2014}, \#24	    &$1277\ km/s$   &initial speed\\
     	&\citeA{Mostl2014}, \#25	    &$1369\ km/s$   &$6.9$ $R_{\circ}$ $f=1$\\
     	&\citeA{Gopalswamy2013}, \#20	&$1210\ km/s$   &COR2, STEREO B \\
     	&\citeA{Sachdeva2017}, \#27	    &$1248\ km/s$   &initial speed at $4.4$ $R_{\circ}$\\
\hline
$28$	&this study                     &$740\ km/s$    &up to $39$ $R_{\circ}$\\
    	&\citeA{Sachdeva2017}, \#28     &$1305\ km/s$   &initial speed at $6.7$ $R_{\circ}$\\
\hline
$29$	&this study                     &$692\ km/s$    &up to $77$ $R_{\circ}$\\
    	&\citeA{Barnard2017}, \#3       &$698\ km/s$    &\\
     	&\citeA{Sachdeva2017}, \#29     &$790\ km/s$    &initial speed at $31.1$ $R_{\circ}$\\
\hline
$30$	&this study                     &$431\ km/s$    &up to $45$ $R_{\circ}$\\
    	&\citeA{Sachdeva2017}, \#30&	$570\ km/s$     &initial speed at $36.9$ $R_{\circ}$\\
\hline
$33$	&this study                     &$765\ km/s$    &$41$ $R_{\circ}$\\
    	&\citeA{Sachdeva2017}, \#33&	$1504\ km/s$    &initial speed at $5.9$ $R_{\circ}$\\
\hline
$34$	&this study	                    &$764\ km/s$    &$39$ $R_{\circ}$\\
    	&\citeA{Sachdeva2017}, \#34&	$1115\ km/s$    &initial speed at $5.9$ $R_{\circ}$\\
 \hline
\end{tabular}}
\end{table}

\subsection{Comparison with results from Heliospheric Imager Geometrical Catalogue (HIGeoCAT)}
\label{sec:comparing_with_helcats}

The HELCATS Heliospheric Imager Geometrical Catalogue (HIGeoCAT) \cite{Barnes2019} reports speeds derived from STEREO/HI observations without considering observations from coronagraphs. Since this catalog covers most events studied here, we compare its results with those derived here.

As introduced in Section \ref{sec:intro}, the reported speeds in the HIGeoCAT catalog are derived using three \textit{single-spacecraft\/} geometric  models, namely f-$\phi$, HM, and SEEF. By comparing the timing of each event in HIGeoCAT with our height-time points, we identify the HIGeoCAT event that corresponds to our event. This was done for both viewpoints (STEREO-A and STEREO-B). We did not find any CME in HIGeoCAT STEREO-A event list corresponding to our event \#25. For the remaining 13 events, we compare the HiGeoCAT speeds  with $v_{med}$ (mentioned in Table \ref{tab:elevoparams}). Given the 3 fits and the two viewpoints available, each CME speed derived here can be compared to 6 different speeds from HIGeoCAT. \protect These results are all summarized in Table \ref{tab:comparing_helcats}.

Significant agreement between HIGeoCAT reported speeds and ours is not expected. First, HIGeoCAT speeds are calculated using both HI-1 and HI-2 observations while we only use HI-1. This catalog reports linear speeds derived over both heliospheric imagers FOVs and CMEs can be accelerated or decelerated while within HI-2 FOV. Second, we derive the CME speed in the ecliptic plane while the HIGeoCAT speeds are not necessarily measured in the ecliptic plane; the speeds are derived at the position angle of the CME apex. For example, for the event \#3, the speed is derived at more than $10^{\circ}$ above the ecliptic. Third, the HIGeoCAT speeds are calculated independently for each viewpoint while our speeds are obtained considering both. (Although HIGeoCAT does associate the observations of each event from both STEREO viewpoints, there is no reported speed obtained considering the combined dual-viewpoint.)

As a result of using the viewpoints independently from one another,  the reported speed values are different for any given event  (see Table \ref{tab:comparing_helcats}). For example, the speeds reported for event \#2 computed using the HM method are $962\pm24$ km/s from STEREO-A and $1368\pm213\ km/s$ from STEREO-B. Among the 13 events compared here, the median absolute difference between the speeds derived using HM for the two STEREO viewpoints is $140\ km/s$ with standard deviation of $256\ km/s$. The differences are similar for the derived speeds based on f-$\phi$ and SSEF. The median absolute difference between $v_{med}$ and the speed derived using f-$\phi$ on STEREO-A (STEREO-B) is $53\pm140\ km/s$ ($68\pm84\ km/s$).  Comparing the results derived using f-$\phi$ with observations from STEREO-A and those from STEREO-B, the median absolute difference found is $96\pm109\ km/s$. We understand that these differences are acceptable considering the several differences between the assumptions behind our methodology and those from the f-$\phi$ method. 

Some events, though, exhibit much larger differences depending on the methodology used, both between the different fittings used on HIGeoCAT and between our speeds and the HIGeoCAT speeds. This is the case for events \#27 and \#28, which show differences higher than $500\ km/s$. For the remaining events, the differences are mostly below $100\ km/s$. The  reason for this large difference is not obvious and needs to be investigated.

   \begin{sidewaystable}
  \caption{Comparison of speeds ($km/s$) calculated in this study ($v_{med}$ second column) with HELCATs (3rd to 7th columns, from left to right). \say{A} or \say{B} before the acronym of each method the viewpoint used (STEREO-A or STEREO-B, respectively). The third column, from right to left, indicates the difference between SSEF derived using STEREO-A and SSEF derived using STEREO-B. The remaining two columns indicate the difference between SSEF and $v_{med}$.}
 \label{tab:comparing_helcats}
 \scalebox{0.75}{
\begin{tabular}{c c c c c c c c c c c}
 \hline

\# & $v_{med}$	& A f-$\phi$	&A SSEF &	A HM	&B f-$\phi$ &	B SSEF &	B HM		& A SSEF - B SSEF &	A SSEF - $v_{med}$ &	B SSEF - $v_{med}$\\
 \hline
$2$     &$846$	    &	$889\pm17$  &	$927\pm20$  &	$962\pm24$  &	$1149\pm106$&	$1248\pm150$&	$1368\pm213$	& -321  &	81  &	402\\
$3$     &$490$      &	$506\pm10$  &	$514\pm12$  &	$520\pm14$  &	$558\pm33$  &	$581\pm50$  &	$605\pm70$      & -67   &   24  &	91\\
$9$     &$465$      &	$412\pm1$   &	$415\pm1$   &	$416\pm2$   &	$508\pm6$   &	$523\pm8$   &	$535\pm11$      &-108   &	-50 &	58\\
$11$    &$448$      &	$409\pm6$   &	$440\pm7$   &	$475\pm9$   &	$425\pm14$  &	$434\pm18$  &	$442\pm22$      &	6   &	-8  &	-14\\
$13$    &$769$	    &	$805\pm26$  &	$947\pm40$  &	$1164\pm69$ &	$775\pm10$  &	$796\pm7$   &	$806\pm7$       &151    &   178	&   27\\
$17$    &$605$	    &	$696\pm22$  &	$745\pm37$  &	$800\pm55$  &	$501\pm9$   &	$563\pm17$  &	$645\pm32$      &182	&   140	&   -42\\
$25$    &$446$	    &	-	        &   -           &   -           &   $585\pm8$   &	$610\pm12$  &	$634\pm18$      & -	    &  -    & -  \\
$26$    &$741$      &	$791\pm36$  &	$877\pm51$  &	$983\pm74$  &	$877\pm16$  &	$942\pm21$  &	$1017\pm28$     &	-65 &	136	&   201\\
$27$    &$743$      &	$1285\pm84$ &	$1573\pm139$& $2070\pm280$	&.  $930\pm65$  &	$1036\pm84$ &	$1172\pm113$    &	537 &	830 &	293\\
$28$    &$740$      &	$921\pm40$  &	$1093\pm57$ &	$1369\pm90$ &	$653\pm24$  &	$728\pm38$  &	$827\pm60$      &	365 &	353 &	-12\\
$29$    &$692$      &	$620\pm14$  &	$700\pm18$  &	$807\pm26$  &	$675\pm30$  &	$717\pm47$  &	$765\pm68$      &	-17 &	8   &	25\\
$30$    &$431$	    &	$396\pm11$  &	$430\pm16$  &	$471\pm22$  &	$485\pm15$  &	$540\pm23$  &	$611\pm35$      &	-110&	-1  &	109\\
$33$    &$765$	    &	$731\pm12$  &	$807\pm13$  &	$902\pm14$  &	$752\pm15$  &	$840\pm24$  &	$954\pm38$      &	-33 &	42  &   75\\
$34$    &$764$	    &	$590\pm2$   &	$682\pm1$   &	$810\pm4$   &	$687\pm9$   &	$809\pm10$  &	$986\pm13$      &	-127&	-82 &   45\\
 \hline
\end{tabular}}
\end{sidewaystable}

\subsection{The CME Time-of-Arrival (ToA) errors}
\label{sec:results_toa_erros}
\begin{figure}
 \noindent\includegraphics[width=\textwidth]{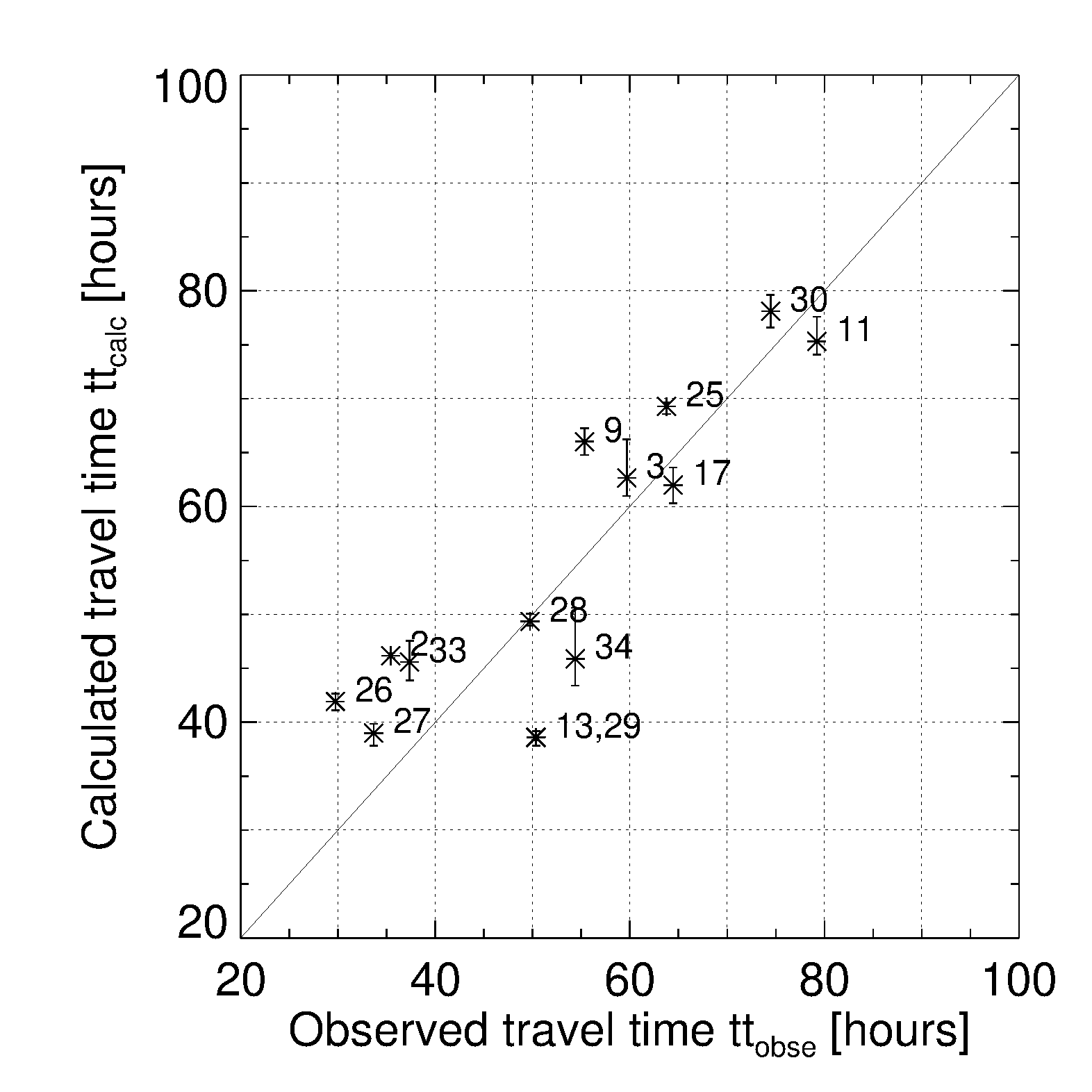}
\caption{The calculated and observed travel time (from the last observation on HI-1 FOV until L1). The labels correspond to the CME IDs in Table \ref{tab:events}. The  line indicates the points where the model and observed travel times are identical, i.e., the ToA error is zero ($\delta t=0$).}
\label{fig:tt}
\end{figure}

In this section we compare the CME travel time from its last HI-1 observation point $s_0$ up to 1 au calculated using the ElCon model and the aerodynamic drag model ($tt_{calc}$) with the actual travel time ($tt_{obse}$). The latter is the time difference between the first ICME observation and the last CME observation at $s_0$. The results are shown in Figure \ref{fig:tt}.

The instant of the CME arrival at Earth is clearly identified from in situ observations for all events studied here. All events are preceded by a clear discontinuity in the magnetic field and solar wind parameters (solar wind speed, density and temperature). For this reason, it is unlikely that the CME ToA errors found here are due to ambiguous determination of ICME arrival time. Since $tt_{obse}$ is not expected to be a source of errors, we focus this study on sources of errors associated to $tt_{calc}$.

For the 14 CME-ICMEs pairs studied here, the ToA error mean value is $1.6\pm8.0$ hours and the mean absolute error (MAE) is $6.9\pm3.9$ hours. The Pearson correlation coefficient found when we compare $tt_{calc}$ and $t_{obs}$ is $0.85$. 

\protect The MAE found here is one of the lowest MAEs according to the \citeA{Vourlidas2019} review of CME ToA that considers more than 20 studies of ToA error. Among the previous studies, 7 of them consider the drag force (the majority using empirical values of $\gamma$ rather than from the model we adopted here); only one study adopted the ElCon model (although without using simultaneous observations from STEREO). \protect The low MAE is probably not surprising given our small sample size (only 14 CMEs). It does cover about a third of the total possible sample (see Section \ref{sec:list}), over 4 out of the 8 years of STEREO-B, and over the rise to solar maximum. Other studies using drag-based model included up to 34 events and other references about empirical methods have more than 200 CMEs in their sample. The main reasons for the small data set are our rather strong selection criteria. As mentioned in Section \ref{sec:list}, we require simultaneous observations from both HI-1s, events well-separated in time/space and reliable CME-ICME identifications. 

 \begin{sidewaystable}
 \centering
  \caption{Calculated and observed CME travel time and speed between end HI-1 FOV and L1.}
 \label{tab:tt}
 \scalebox{0.85}{
\begin{tabular}{c c c c c c c c c c c c c c c c c}
 \hline
 ID& first observation & last tracking &  Arrival at $1\ au$ & $tt_{calc}$& $tt_{obse}$& $\delta t$ & $v_{init}$& $v_{final}$ & $v_{final}^{+}$ & $v_{final}^{-}$& $v_{ICME}$& $v_{SW@1au}$&$n_{SW@1au}$\\
  & (UT)   &   (UT) &  (UT)        & $[h]$& $[h]$& $[h]$ & $[km/s]$& $[km/s]$ & $[km/s]$ & $[km/s]$& $[km/s]$& $[km/s]$&$[cm^{-3}]$\\ 
 \hline
$  2$&03-Apr-2010 10:33:58&03-Apr-2010 20:29:21&05-Apr-2010 07:55:00& 46& 35& 11&863&662&  663&  660&  704&  620&    4&  \\
$  3$&08-Apr-2010 04:54:07&09-Apr-2010 00:39:22&11-Apr-2010 12:20:00& 63& 60&  3&476&466&  478&  443&  431&  401&    5&  \\
$  9$&15-Feb-2011 02:24:05&15-Feb-2011 18:29:34&18-Feb-2011 01:50:00& 66& 55& 11&465&456&  464&  448&  510&  378&    5&  \\
$ 11$&25-Mar-2011 08:48:25&26-Mar-2011 07:59:25&29-Mar-2011 15:12:00& 75& 79& -4&440&436&  443&  423&  379&  346&    5&  \\
$ 13$&14-Jun-2011 06:12:05&14-Jun-2011 23:49:28&17-Jun-2011 02:09:00& 39& 50&-11&770&756&  761&  752&  521&  448&    5&  \\
$ 17$&14-Sep-2011 00:00:05&14-Sep-2011 10:29:53&17-Sep-2011 02:57:00& 62& 64& -2&586&558&  572&  546&  480&  455&    5&  \\
$ 25$&19-Apr-2012 15:12:09&20-Apr-2012 10:29:25&23-Apr-2012 02:15:00& 69& 64&  5&448&445&  450&  443&  384&  340&    5&  \\
$ 26$&14-Jun-2012 14:12:07&15-Jun-2012 03:19:22&16-Jun-2012 09:03:00& 42& 30& 12&757&676&  687&  666&  478&  408&   10&  \\
$ 27$&12-Jul-2012 16:48:05&13-Jul-2012 07:59:27&14-Jul-2012 17:39:00& 39& 34&  5&752&701&  719&  688&  606&  500&    9&  \\
$ 28$&28-Sep-2012 00:12:05&28-Sep-2012 08:29:50&30-Sep-2012 10:14:00& 49& 50& -1&733&639&  643&  630&  310&  279&    5&  \\
$ 29$&05-Oct-2012 02:48:05&06-Oct-2012 01:49:52&08-Oct-2012 04:12:00& 39& 50&-11&697&659&  672&  650&  387&  323&    5&  \\
$ 30$&27-Oct-2012 16:48:05&28-Oct-2012 11:59:57&31-Oct-2012 14:28:00& 78& 74&  4&431&405&  412&  399&  353&  279&    5&  \\
$ 33$&15-Mar-2013 07:12:05&15-Mar-2013 15:59:43&17-Mar-2013 05:21:00& 46& 37&  9&735&721&  747&  693&  679&  536&    5&  \\
$ 34$&11-Apr-2013 07:24:06&11-Apr-2013 15:49:33&13-Apr-2013 22:13:00& 46& 54& -8&737&728&  769&  661&  491&  421&    2&  \\
  \end{tabular}}
  \end{sidewaystable}

The relatively low ToA error found here suggests that events observed in close timing with others, which were discarded by our criteria, will likely  increase ToA errors. The criteria adopted here limit the use of this methodology on routine space weather applications that that can not perform event selection and need to measure all CMEs. Carefully selected and investigated event samples can help isolate physical effects during CME propagation from analysis errors (e.g. front identification) and hence help improve our physical understanding of these events and eventually space weather forecasting.

Our results, along with the \citeA{Colaninno2013} and \citeA{Rollett2016} results, suggest that the estimation of the ToA using HI-1 measurements could result in a more accurate estimation (i.e., smaller error) than in those cases based on coronagraph observations, at least for fast CME events. This conclusion is based on studies with a rather small list of event and is therefore subject to verification with more extensive data sets. In addition, many of the methods rely on coronagraph measurements of CME width, mass, and other properties. Hence, it is more likely that approaches built upon as extensive height-time measurements as possible, will be more fruitful in reducing the errors of the ToA estimation.

\subsection{ The CME Speed-on-Arrival (SoA) and its error}
\label{sec:Soa_and_error}
Here, we compare the CME SoA at $1\ au$ derived from our drag-based model with the corresponding in situ observed ICME speed ($v_{ICME}$). 

\protect The in situ ICME speed, $v_{ICME}$, is derived here as the average proton speed observed in situ during the sheath period, not during the whole time period comprised by the passage of the ICME. The use of this time period is motivated by the higher proton density of the sheath feature (this density is a common physical parameter to both the in situ and imaging instruments). After the sheath, a region with lower density and smooth magnetic field (the magnetic cloud) is observed, which does not correspond to the front we identify on HI-1 observations. Heliospheric imagers identify the compression region (sheath) developed around the ejecta, and not necessarily the magnetic cloud. The in situ data used here comes from the OMNI database and consists of merged observations from the ACE and the Wind spacecraft \cite{King2005}.

In this study, we calculated the SoA using 3 different initial speeds in the drag model for each CME ($v_{med}$, $v_{min}$ and $v_{max}$). The SoA derived  are labelled $v_{final}$, $v_{final}^{-}$ and $v_{final}^{+}$, respectively (Table \ref{tab:tt}). The difference between $v_{med}$, $v_{min}$ and $v_{max}$ comes from the multiple visual CME identification in the J-map \protect . Due to the subjective CME identification, every visual inspection led to slightly different elongation-time profile since the specific point identified changes (see details in Section \ref{sec:jmap}). We found that $|v_{max}-v_{min}|$  is $<\ 50\ km\ s^{-1}$ for all events, except for \#34 ($113 \ km\ s^{-1}$). The mean value of $v_{final}^{+} \ -\ v_{final}^{-}$ is $27$ $km\ s^{-1}$. \protect Compared to the typical CME speed error of $\approx 100\ km\ s^{-1}$ found by \citeA{Mierla2010} when comparing several methodologies with observations from SECCHI coronagraphs, the SoA uncertainty caused by multiple visual CME identification on J-maps is quite low.

This result suggests that the difference in the visual selection of features on a J-map (which is responsible for the difference in the initial speed used on the drag model) lead to a minor differences in the SoA. An example of CME speeds as a function of position calculated using both $v_{min}$ and $v_{max}$ is shown in Figure \ref{fig:dragexa} (second panel, from top). They are represented by the blue and red lines, respectively. In this example the difference between $v_{final}^{-}$ and $v_{final}^{+}$ is $31\ km\ s^{-1}$.

\begin{figure}
\includegraphics[width=\textwidth]{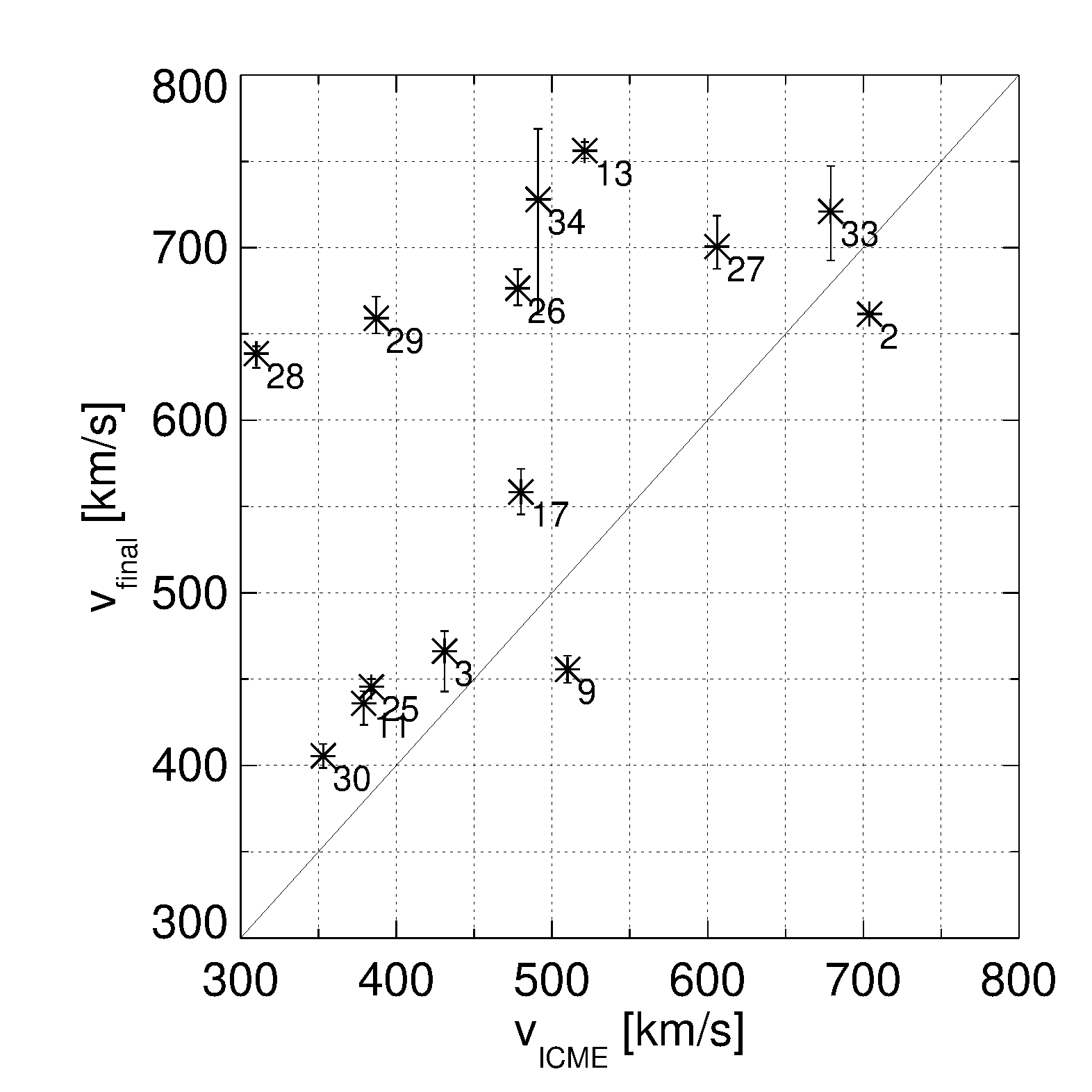}
\caption{The CME speed calculated at $1\ au$ using the drag model ($v_{final}$) compared to the in situ ICME speed ($v_{ICME}$). The labels correspond to the CME IDs in Table \ref{tab:events}. The line represents the region with $\delta v=0$, i.e., the position a given event would be located if it had null SoA error.}
\label{fig:v_observed_calculated}
\end{figure}

The distribution of $v_{final}$ versus $v_{ICME}$ is shown in Figure \ref{fig:v_observed_calculated}. The error bars shown in the plot are defined by $v_{final}^{+}$ and $v_{final}^{-}$. The  CME SoA error ($\delta v = v_{final}\ -\ v_{ICME}$) is $114\pm119$  $km\ s^{-1}$ and the SoA MAE is $117\pm102$ $km\ s^{-1}$. It is clear that our SoA are higher than the observed ICME speeds. The Pearson correlation coefficient between $v_{final}$ and $v_{ICME}$ is $0.53$, lower than the correlation found comparing observed and calculated travel times ($0.85$).

The SoA error is not reduced significantly when we consider $v_{final}^{+}$ and $v_{final}^{-}$. This indicates that the error in the initial CME speed, estimated via multiple visual identifications on the J-maps, cannot explain the majority of the SoA error. We offer a few plausible additional sources of error below: 
\begin{itemize}
    \item an error in the initial CME speed ($v_{init}$) due to the ElCon model and its assumptions, such as linear speed and fixed direction of propagation. The goodness of the fit  $\sigma$ (Table \ref{tab:elevoparams}) is quite low for most events ($<0.01\ au$), $\sigma$ being higher for event \#27. We found that SoA and ToA errors are not correlated with $\sigma$ or parameters from the elliptical model such as half-width $\lambda$ nor aspect ratio $f$. However, we do not have estimates of ElCon model contribution on the SoA error;
 
    \item the $v_{ICME}$ does not correspond to the CME front speed precisely. This could happens because $v_{ICME}$ is measured in situ from observations of the solar wind particles around the observing spacecraft. It is widely known from in situ observations that this speed is highly variable over time and position, even within periods and dimensions that are typically associated to ICMEs \cite<see, e.g., >{Richardson2010}. On the other hand, $v_{final}$ is a parameter that describes the CME front as a whole, which has spatial dimensions that are orders of magnitude larger than the region observed in situ by a spacecraft;
 
    \item an incomplete or incorrect description of the forces that affect the CME propagation from $s_0$ to $1 \ au$. This can impact the CME dynamics in all along its propagation from the Sun to the Earth.
\end{itemize}

The SoA MAE was compared with results from 5 other studies, as shown in Table~1 of \citeA{Vourlidas2019}. Our results are similar except the much smaller SoA in \citeA{Rollett2016} ($16\pm53$ $km\ s^{-1}$). 

We identified that 5 events studied here (\# 13, \#26, \#28, \#29 and \#34) have SoA absolute error ($|\delta v| =|v_{final}-v_{ICME}|$) higher than $198$ $km\ s^{-1}$ while the remaining are lower than $95\ km\ s^{-1}$ (this can be seen clearly in Figure \ref{fig:v_observed_calculated}).
We tried to identify any trend between the $|\delta v|$ and input parameters used in the drag model, particularly those that change between events. We could not find, though, any trend between $|\delta v|$ and CME mass, width, background solar wind speed in the corona and 1 au nor solar wind density at 1 au. 

One common point among events with higher $|\delta v|$ is that the CME initial speed is between $\sim600$ $km\ s^{-1}$ and $\sim 800\ km\ s^{-1}$. The opposite is not true, however. Some CMEs with initial speed in the same range have $|\delta v|$ among the lowest values ($<\ 95\ km\ s^{-1}$). This result suggests that the CME propagation modeling used here (ElCon and our drag force description) do not lead to higher SoA error for any particular range of CME initial speeds. 

\section{Discussion}

In this Section, we focus on the ToA and SoA errors and their possible sources. Namely, we examine  the influence of the background solar wind conditions (Section \ref{sec:toa_and_soa_different_background}), the drag model assumptions (Section \ref{sec:contributionSoA}), and the effects of extending tracking further into the heliosphere (Section \ref{sec:tracking_cme_fruter_toa_soa_erros}). We also investigate the ToA error when we completely remove the drag force and consider a CME propagating with constant speed up to the Earth (Section \protect \ref{sec:contributionToA}). Our findings are summarized in Table~\ref{tab:comparing_errors}. The first column indicates, shortly, the possible sources of error. Details about each case are explained in Sections \ref{sec:Soa_and_error} to \ref{sec:tracking_cme_fruter_toa_soa_erros}. The second column indicates the variable associated to the corresponding error source. In some cases, we compare the variable with the SoA and ToA to evaluate any correlation between them. The last column (right) states whether the corresponding plausible sources of error are likely to result in SoA and ToA errors comparable to those found in our work.
 \begin{sidewaystable}
    \centering
  \caption{Summary parameters used on the drag force estimation and their expected contribution on ToA and SoA errors.}
 \label{tab:comparing_errors}
\begin{tabular}{c c c c }
 \hline
Possible source of error &  Variable & \parbox[t]{3cm}{\centering Correlation with SoA and ToA errors} & \parbox[t]{5cm}{\centering Magnitude large enough to explain SoA and ToA errors found here} \\
 \hline
Speed error due to visual CME identification on J-maps &     $|(v_{max}-v_{min})|$ & No & No\\
Residue of the elliptical front determination &  $\sigma$ & No & Unknown\\
Background solar wind density at $1\ au$& $n_{SW@1AU}$ & No & No\\
Background solar wind speed at $1\ au$ & $v_{SW@1AU}$ & No & No \\
Position of last CME observation on HI1-FOV & $s_0$ & ToA (low) & Unknown \\
CME cross section area  & $A_{CME}$ & No & Yes\\
CME mass & $m_{CME}$ & No & Yes\\
Incomplete or incorrect force description  & $a$ & Unknown & Unknown\\
\end{tabular}
\end{sidewaystable}

\subsection{CME ToA and SoA errors during different background solar wind conditions}
\label{sec:toa_and_soa_different_background}

Now we examine the effects of the background solar wind conditions, such as proton density or speed, on the drag model and by extension on ToA and SoA.

The drag force depends on the difference between CME and background solar wind speed. Our events occur over a diverse range of 1 au solar wind speeds, $v_{sw@1au}$, as listed in the second column of Table \ref{tab:tt}, from right to left. Since $v_{sw@1au}$ is used to extrapolate the solar wind speed to $s_0$, it affects the drag force used in the model. 

For all our events, $v_{sw}$ is lower than $v_{CME}$ at the first height of application of the drag force ($s_0$) and, as a result, the drag force produces deceleration. In 11 of the 14 events  $v_{sw@1au}$ was lower than $500\ km/s$. The highest value of $v_{sw@1au}$ was observed in CME $\#2$: $620\ km/s$.  In some events (such as $\#28$ and $\#30$), the solar wind speed is quite low $v_{sw@1au} = 279\ km/s$. We do not find any trend between $v_{sw@1au}$ and ToA or SoA errors. This can be due to either the drag force description we use is insensitive to background solar wind speed or the $\delta t$ and $\delta v$ originate from sources other than the drag force, such as errors in the determination of initial CME speed, direction of propagation or position.

Another solar wind parameter of the drag model is the background solar wind density at $1\ au$, which is used in the drag force calculation to estimate the solar wind density along the CME path. For only two events ($\#26$ and $\#27$) the solar wind density is higher (by about a factor of 2) than the average for quiet periods ($n_{SW@1au}= 5 \ cm^{-3}$). Again, we find no trend between the background solar wind density and $\delta t$ or $\delta v$. This result suggests that the drag model estimates are insensitive to the details of the background solar wind density, at least, for the range of values used here.

However, we note that the background solar wind density and speeds considered here are just model-based values. The actual heliospheric conditions may be very different due, for example, to the existence of transients such as, other CMEs or stream interaction regions (SIRs). Although we tried to exclude periods with multiple CMEs in the HI-1 FOV (see Section \ref{sec:list}), we did not check for the existence of upstream CMEs or SIRs.

In particular, we notice that at least one of the events studied here (\#9) is preceded by some CMEs observed on COR2 (see details in \citeA{Gopalswamy2013}). This could at least partially explain the $\delta t = -11 h$ found for this event.

\subsection{The effect of drag on the CME SoA}
\label{sec:contributionSoA}

The CME speed variation ($\Delta v = v_{final} - v_{init}$) in the entire range we applied the drag force (from $s_0$ up to $1\ au$)  is $> 100\ km\ s^{-1} $ for event $\#2$ and $\Delta v\ >\ 50\ km\ s^{-1}$ for the following events: $ \#26$,  $\#27$ and $\#28$. On the remaining 10 events, $0\ <\ \Delta v\ <\ 50\ km\ s^{-1}$.

Since the typical error of CME speed in coronagraph observations is around $100\ km\ s^{-1}$ \cite{Mierla2010}, we conclude that the contribution of drag on the SoA is small and within the error range of the CME speed observations, at least for the events considered here. Our results agree with \citeA{Sachdeva2015} who found that the drag force is minimum at distances above  15-50 solar radii for slow CMEs since they propagate almost at constant speeds after that range. 

A second point is that all events with $\Delta v\ > 30\ km\ s^{-1}$ have $v_{init}\ ~>$ $733$ $km\ s^{-1}$ but some events  with $v_{init} >$ $733$ $km\ s^{-1}$ ($\#13, \#33$ and $\#34 $) have $\Delta v\ <\ 30\ km\ s^{-1}$. This result illustrates that although the drag force absolute value is frequently higher for high-speed CMEs, factors other than the CME initial speed strongly affect some events.

\subsection{The effect of drag on the CME ToA}
\label{sec:contributionToA}

To assess the effect of  drag  in the estimation of the ToA, we repeated the ToA calculation without the drag force. This corresponds to a very simplified model consisting of a CME propagating from $s_0$  to $1\ au$ with constant speed, which equals $v_{init}$.

The ToA mean error considering no drag force is $-0.4\pm7.4\ h$ and ToA mean absolute error is $6.1\pm3.9\ h$. Comparing these values to the results found using the drag force, we can see that they are identical within the error range.  Therefore, the contribution of the drag force is at most at the same level of magnitude than other unknown reasons that drive the ToA error. As discussed in Section \ref{sec:contributionSoA}, results from previous studies using the same drag force model suggest that the effect of this force is not very significant at the heliocentric distances range where $s_0$ typically lies.

This does not mean that drag is negligible for CME propagation studies. The drag force is likely stronger closer to the Sun than at the locations studied here ($s_0$) because the solar wind speed is lower and, at least for fast CMEs, the CME speed is higher. 

\subsection{Does Tracking the CME further in the HI-1 FOV reduces ToA and SoA errors?}
\label{sec:tracking_cme_fruter_toa_soa_erros}
The drag model does not start at the same position for all 14 events. Each CME is tracked until $s_0$, which is the last point where it is clearly observed in HI-1 FOV. Then, drag is applied from this point up to $1\ au$, as explained in Section \ref{sec:model}. CMEs with lower $s_0$ have their speed, direction of propagation and morphological parameters (such as angular width in the ecliptic plane and elliptical aspect ratio) derived closer to the Sun.

Within our limited 14-event sample, there is no correlation between the SoA absolute error $|\delta v| = |v_{final}-v_{ICME}|$ and $s_0$ (the Pearson correlation coefficient is $0.16$). CMEs with higher $s_0$ exhibit, however, a tendency toward higher absolute ToA errors $|\delta t|$ (in this case the correlation coefficient is $0.46$).

This trend could arise from the following considerations: (i) as a CME moves away from the Sun, its brightness decreases in the HI-1 FOV and hence the identification of its front becomes more ambiguous; and (ii) errors associated with the ElCon assumptions about CME kinematics. As described in Section \ref{sec:elliptical}, we are assuming linear speed and fixed direction of propagation for each CME up to $s_0$, beyond that point we use ElCon to derive the CME parameters. Beyond $s_0$, a free parameter for acceleration is included but the direction propagation is still assumed to be constant.

The last consideration to explain the trend observed for higher $|\delta t|$ is also pointed out on \citeA{Barnard2017}. The authors observed unrealistic acceleration in regions close to the outer side of the HI-1 FOV, mainly after typical values of $s_0$. The same study also found unrealistic accelerations when other methods with constraints in the direction of propagation were used, such as harmonic mean and self-similar expansion. In this way, the results from \citeA{Barnard2017} seem to support hypothesis (ii) as the explanation for a tendency toward a higher $|\delta t|$ for the events studied here with higher $s_0$.  

 \section{Summary and Conclusions}
 \label{sec:conclusions}

From an initial list of 38 Earth-directed CMEs in 2010-2013 compiled by \citeA{Sachdeva2017}, we selected 14 events by applying three rather strict criteria: simultaneous observations from both STERO/HI-1 instruments, a clear CME-ICME counterpart identification, and events separated in time to avoid CME-CME interactions. Our objective was to minimize as much as possible the source of errors in the measurements of the CME kinematic parameters and ToA. The arrival time of all 14 events could be unambiguously determined from in-situ observation thanks to a discontinuity clearly observed in both magnetic field and solar wind plasma parameters.

We extracted the kinematics of the events using observations from HI-1, modeled their front using ElCon, and extrapolated both their time-of- and their speed-on- arrival using a drag force model. The modeled CME speed at $1\ au$ was typically higher than the observed ICME speed. This was the case for all events analyzed but one ($\#9$). SoA absolute errors are higher than $198\ km/s$ for 5 events ($\#13$, $\#26$, $\#28$, $\#29$ and $\#34$) and below $120\ km/s$ for the remaining 8 events. This suggests that either the actual initial CME speed was lower than what our measurements suggested or that the deceleration magnitude calculated using the drag-based model studied here was lower than the actual one. The latter seems to be a more likely explanation since excess SoA is a common result in many studies \cite{Vourlidas2019}. 

The resulting ToA absolute errors are below 12 hours when considering all 14 events. Our MAE compares favorably against past studies and is encouraging regarding our approach. However, the results are based on a small number of events and  the methodology may not necessarily lead to lower ToA when applied to more CMEs. We plan to pursue this further by addressing the various issues we identified below.

Sources of ToA and SoA errors can arise in drag force calculation or in the presence of  other unaccounted for forces, such as the Lorentz force. Another source of error may be the assumption of the elliptical conversion model used for the determination of the CME radial position from its elongation, such as fixed direction of propagation and constant speed. Finally, errors on the front identification arise towards the outer FOV of HI-1  as the CME front becomes fainter.

The drag force calculation at any point in the CME trajectory depends on the ambient solar wind density and speed. These conditions can change significantly during CME propagation and unfortunately in-situ observations were available only close to the Earth for the CMEs under study. In this study, both solar wind density and speed were extrapolated using empirical expressions. For this reason, the drag force should be understood as an approximation rather than a precise calculation. More realistic solar wind conditions derived using simulation are out of the scope of the present manuscript and could be part of a future study.

The amplitude of the drag force is stronger close to the Sun when compared to conditions close to the Earth. The reason is twofold: (i) the difference between the solar wind and the CME speeds and (ii) the density profile of the solar wind, which is higher close to the Sun (typically by one order of magnitude at 50 solar radii when compared to L1). 

Deceleration was observed in all 14 events since all had initial speeds higher than the solar wind speed at the starting point of the drag force application. This deceleration is more intense close to the Sun, where the background solar wind speed is also lower. The deceleration reaches values up to $-3.25\ m\ s^{-2}$ close to the Sun and $-0.5\ m\ s^{-2}$ close to the Earth. 

Despite the difficulties to track CMEs in the HI-1 FOV due to the presence of the F-corona and reduced CME brightness, the results suggest that the ToA error is similar to many studies based on coronagraph observations, at least for the events discussed here.

The recently (2018) launched Parker Solar Probe (PSP) Mission \cite{Fox2015a} has an imager instrument with comparable elongation range to the HI-1 used. This imager is the Wide-field Imager for Solar PRobe (WISPR) \cite{Vourlidas2016}. Similar observations will also be performed by the Solar Orbiter Heliospheric Imager (SoloHI) \cite{Howard2019}, onboard the upcoming Solar Orbiter (SO) mission \cite{Muller2013}. 
In this sense, the present study, which relies mostly on observations from heliospheric imagers (using only masses and width derived from coronagraph observations), can be used as a guideline for future studies with the PSP and SO targeted on CME ToA or SoA estimations. We hope the results of CME ToA errors estimated could motivate future studies with similar objectives using observations from WISPR and SoloHI.

\acknowledgments

C.R.B. acknowledges grants \#2014/24711-6 and \#2017/21270-7 from S\~{a}o Paulo Research Foundation (FAPESP). A.V. is supported by NASA grant 80NSSC19K1261. G.S. and C.R.B. acknowledge the support from the NASA STEREO/SECCHI (NNG17PP27I) program. E.E. acknowledges grants \#2018/21657-1 from S\~{a}o Paulo Research Foundation (FAPESP) and \#301883/2019-0 from CNPq/PQ. A.D.L acknowledges grant \#309916/2018-6 from CNPq/PQ.

The Large Angle Spectrometric Coronagraph instrument (LASCO) was constructed by a consortium consisting of the Naval Research Laboratory (Washington DC, USA), the Max Planck Institute for Solar System Research (currently in Gottingen, Germany, formerly known as the Max Planck Institute for Aeronomie in Kathenburg, Lindau, Germany), the Laboratoire d'Astronomie Spatiale (Marseille, France), and the Space Research Group at the University of Birmingham (Birmingham, UK). LASCO is one of a complement of instruments on the Solar Heliospheric Observatory satellite (SOHO) built in an international collaboration between the European Space Agency (ESA) and National Aeronautics and Space Administration (NASA). LASCO data are available for download at \url{https://lasco-www.nrl.navy.mil/index.php?p=content/retrieve/products}.

The SOHO LASCO CME catalog is generated and maintained at the CDAW Data Center by NASA and The Catholic University of America in cooperation with the Naval Research Laboratory.  This catalog is available at \url{https://cdaw.gsfc.nasa.gov/CME_list/}.

The Sun Earth Connection Coronal and Heliospheric Investigation (SECCHI) was produced by an international consortium of the Naval Research Laboratory (USA), Lockheed Martin Solar nd Astrophysics Lab (USA), NASA Goddard Space Flight Center (USA), Rutherford Appleton Laboratory (UK), University of Birmingham (UK), Max Planck Institute for Solar System Research (Germany), Centre Spatiale de Liége (Belgium), Institut d’Optique Theorique et Appliquée (France), and Institut d'Astrophysique Spatiale (France). STEREO data are available for download at \url{https://secchi.nrl.navy.mil/}.

This research has made use of the Solar Wind Experiment (SWE) and Magnetic Field Investigations (MFI) instrument's data onboard WIND. We thank to the Wind team and the NASA/GSFC's Space Physics Data Facility's CDAWeb service to make the data available. Wind data are available from \url{https://cdaweb.sci.gsfc.nasa.gov}. The ICME list compiled from WIND mission observations can be found at \url{https://wind.nasa.gov/ICMEindex.php}. The OMNI data were obtained from the GSFC/SPDF OMNIWeb interface at \url{https://omniweb.gsfc.nasa.gov}.

The HELCATS catalogs are available from the HELCATS website (\url{https://www.helcats-fp7.eu}), HIGeoCat (\url{https://doi.org/10.6084/m9.figshare.5803176.v1}) and HIJoinCat (\url{https://www.helcats-fp7.eu/catalogues/wp2_joincat.html}). 

The codes and script created to perform this study are available in the following repository: \url{https://doi.org/10.7910/DVN/J7SYTO}.

\bibliography{refs_agu_2019}

%
%
%
%
%

\end{document}